\documentclass[twoside,12pt]{article}
\usepackage{epsfig}

\newcommand{\be}{\begin{equation}}
\newcommand{\ee}{\end{equation}}
\newcommand{\bea}{\begin{eqnarray}}
\newcommand{\eea}{\end{eqnarray}}

\usepackage{amsmath}
\usepackage{amsfonts}
\usepackage{amssymb}
\usepackage{amsxtra}
\usepackage{supertabular}
\begin{document}
\title{ \vspace{1cm}% How can Clifford algebra help to understand
  %properties of the second quantized
 % fermions and the corresponding gauge vector and scalar fields\\
% Achievements of {\it spin-charge-family} theory so far, which unifies
% spins, charges and families of fermions, connecting fermions with 
% corresponding vector and scalar gauge fields  \\
The achievements of the {\it spin-charge-family} theory so far %, which unifies
%spines, charges and families of fermions and accordingly also
%tensor, vector, and scalar fields\\
%
%Clifford algebra shows the way to the second quantized fermion, vector and
%scalar gauge fields, unifying spins, charges and families and all the interactions beyond the
%{\it standard model}
}
\author{N.S.\ Manko\v c Bor\v stnik,$^{1}$ \\% H.B.\ Nielsen$^2$\\
$^1$Department of Physics, University of Ljubljana\\
SI-1000 Ljubljana, Slovenia}
% $^2$Niels Bohr Institute, University of Copenhagen\\
% Blegdamsvej 17, Copenhagen\O, Denmark}
\maketitle

\begin{abstract} 
%To be improved, mostly following a part of the review article in Progress in 
% Particle and Nuclear Physics\\
%{\bf A.
% Why we assume only $d$ even, even more $d=2(2n+1)$?
%******************************
%Template   
%
Fifty years ago the {\it standard model} offered an elegant new step
towards understanding elementary fermion and boson fields, making several 
assumptions, suggested by experiments. The assumptions are still waiting 
for an explanation. 
There are many proposals in the literature for the next step. 
The {\it spin-charge-family} theory, proposing a simple starting action in 
$ d\ge (13+1)$-dimensional space with fermions interacting with the gravity 
only (the vielbeins and the two kinds of the spin connection fields), is offering 
the explanation for not only all by the {\it standard model} assumed properties 
of quarks and leptons and antiquarks and antileptons, with the families included, 
of the vectors gauge fields, of the Higgs's scalar and Yukawa couplings,  of  
the appearance of the {\it dark matter}, of  the {\it matter-antimatter 
asymmetry}, making several predictions, but explains as well the second 
quantization postulates for fermions and bosons by using the odd and the even 
Clifford algebra ''basis vectors'' to describe the internal space of  fermions and 
bosons, respectively.  Consequently the single fermion and single boson states 
already anticommute and commute, respectively.
I present in this talk a very short overview of the achievements of the 
{\it spin-charge-family} theory so far, concluding with presenting  not yet solved 
problems, for which the collaborators are very welcome.

\end{abstract}

\section{Introduction}
\label{introduction}

The review article~\cite{nh2021RPPNP} presents a short overview of most of the 
achievements of the {\it spin-charge-family} theory so far. I shall make use of this 
article when presenting my talk. \\

\noindent
Fifty years ago the {\it standard model} offered an elegant new step towards 
understanding elementary fermion and boson fields by postulating: \\
{\bf a.} The existence of massless fermion family members with the  spins 
and charges in the fundamental representation of the groups, {\bf a.i.} the 
quarks as colour triplets and colouress leptons, {\bf a.ii} the left 
handed  members as the weak doublets, the right handed 
weak chargeless members, {\bf a.iii.} the left handed quarks differing from 
the left handed leptons in the hyper charge, {\bf a.iv.} all the right 
handed members differing among themselves in hyper charges, 
{\bf a.v.} antifermions carrying the corresponding anticharges of fermions and 
opposite handedness, {\bf a.vi.} the  families of massless fermions, 
suggested by experiments  and required by the gauge invariance of the boson 
fields (there is no right handed neutrino postulated, 
since it would carry none of the so far observed charges, and correspondingly 
there is also no left handed antineutrino allowed in the {\it standard model}).\\ 
 {\bf b.} The existence of massless vector gauge fields to the observed 
charges of quarks and leptons, carrying charges in the adjoint representations 
of the corresponding charged groups and manifesting the gauge invariance.\\
 {\bf c.}  The existence of the massive weak doublet scalar higgs, {\bf c.i.}
 carrying the weak charge $\pm \frac{1}{2}$  and the hyper charge 
 $\mp \frac{1}{2}$ (as it would be in the fundamental representation of the
 two groups), {\bf c.ii.} gaining at some step of the expanding universe the 
 nonzero vacuum expectation value, {\bf c.iii.} breaking the weak and the 
 hyper charge and correspondingly breaking the mass protection, {\bf c.iv.}
taking care of the properties of fermions and of the weak bosons masses, 
 {\bf c.v.} as well as the existence of the Yukawa couplings. \\
{\bf d.} The presentation of fermions and bosons as second quantized fields.\\
 {\bf e.} The gravitational field in $d=(3+1)$ as independent gauge field.
 (The {\it standard model} is defined  without gravity in
order that it be renomalizable, but yet the standard model particles are ''allowed''
to couple to gravity in the ''minimal'' way.)
\\
 
 %\noindent
 The {\it standard model} assumptions have been experimentally confirmed 
without raising any severe doubts so far, except for some few and possibly statistically 
caused anomalies~\footnote{%
I think here on the improved {\it standard model}, 
in which neutrinos have non-zero masses, and the model has no ambition to explain
severe cosmological problems.}, but also by offering no explanations for 
the assumptions.  
 The last among the fields postulated by the {\it standard model},  the 
scalar higgs, was detected in June 2012,  the gravitational waves 
 were detected in February 2016.
 
The {\it standard model}  has in the literature several  explanations, mostly 
with many new not explained assumptions. The most  popular seem to be 
the grand unifying theories~\cite{Geor,FritzMin,PatiSal,GeorGlas,Cho,ChoFreu,%
Zee,SalStra,DaeSalStra,Mec,HorPalCraSch,Asaka,ChaSla,Jackiw,Ant,Ramond,Horawa}. 
At least $SO(10)$-unifying theories offer the explanation for the postulates from 
{\bf a.i.} to {\bf a.iv}, partly 
to {\bf b.} by assuming that to all the ''fermion'' charges there exist the 
corresponding vector gauge fields --- but does not explain the  assumptions 
{\bf a.v.} up to {\bf a.vi.}, {\bf c.} and {\bf d.}, and does not connect gravity 
with gauge vector and scalar fields.\\

In a long series of works with collaborators~(\cite{norma92,norma93,norma95,%
norma94,IARD2016,IARD2020,n2014matterantimatter,nd2017,n2012scalars,%
JMP2013,nh2017,normaJMP2015,nh2018} and the references therein), we have 
found the phenomenological success with the model named % by N.S.M.B. 
the {\it spin-charge-family}  theory,  with fermions, the internal space of which 
is described with the Clifford algebra of all linear superposition
 of odd products of $\gamma^a$'s in $d=(13+1)$, 
%Sect.~\ref{internalspace},
%(may be  with d infinity), 
interacting with only gravity~(\cite{IARD2020} and references therein). %Sect.~\ref{fermionandgravitySCFT}. 
The spins of fermions from
 higher dimensions, $d>(3+1)$, manifest 
in $d=(3+1)$ as charges of the {\it standard model}, %Sect.~\ref{internalspace},
gravity in higher dimensions manifest as the {\it standard model} gauge vector 
fields as well as the Higgs's scalar and Yukawa 
couplings~%Sect.~\ref{fermionandgravitySCFT}~
\cite{normaJMP2015,nd2017}.

Let be added that one irreducible representation of $SO(13,1)$ contains, 
if looked from the point of view of $d=(3+1)$,
%$2^{\frac{d}{2}-1}$ 
all the quarks and leptons and antiquarks and antileptons and just with the
properties, required by the {\it standard model}, including the relation between 
quarks and leptons and handedness  and antiquarks and antileptons of the 
opposite handedness, as can be reed in Table~5 of App. D, appearing in the 
contribution of the same author in this Proceedings~\cite{norma2021SQFB}. 
%the contribution in the\ref{Table so13+1.}. %, pointed out above. 
%Of course, one needs  the breaks of symmetry from $SO(13,1)$ to
% $SO(7,1)\times SU(3)\times U(1)$ with the condensate. Proof is needed 
% how the condensate breaks the symmetry dynamically.

All that in the {\it standard model} had to be assumed (extremely effective
 "read" from experiments and also from the theoretical investigations) in the 
{\it spin-charge-family} theory appear as a possibility %(degree of freedom) 
from the starting simple action, Eq.~(\ref{wholeaction}), and from the 
assumption that the internal space of fermions are described by the odd 
Clifford algebra objects. 

One can reed in my second contribution to this Proceedings~\cite{norma2021SQFB}
that the description of the internal space of fermions with the odd Clifford algebra 
operators $\gamma^a$'s offers the explanation for the observed quantum numbers 
of quarks and leptons and antiquarks and antileptons while unifying spin, 
handedness, charges and families. The ''basis vectors'' which are 
superposition of odd products of  operators $\gamma^a$'s, 
appear in irreducible representations which differ in the quantum 
numbers determined by $\tilde{\gamma}^a$'s.

The simple starting action of the {\it spin-charge-family} theory offers the 
explanation for not only the properties of quarks and leptons and antiquarks 
and antileptons, but also for  the vector gauge fields, scalar gauge fields, which 
represent higgs and explain the Yukawa couplings, and for the scalars, which 
cause matter/antimatter asymmetry, the  proton decay, while the appearance 
of the dark matter is explained by the appearance of two groups of the
decoupled families.

It appears, as it  is explained in my second contribution to this 
Proceedings~\cite{norma2021SQFB}, 
that the description of the internal space of bosons fields (the gauge fields
of the fermion fields described by the Clifford odd ''basis vectors'') with the 
Clifford even ''basis vectors'' explains the commutativity and the properties
of the second quantized boson fields, as the description of the internal space 
of fermion fields with the Clifford odd ''basis vectors'' explains the 
anticommutativity and the properties of the second quantized fermion fields. 

The description of fermions and bosons with the Clifford odd and Clifford even 
''basis vectors'', respectively, makes fermions appearing in families, while bosons 
do not. Both kinds of ''basis vectors'' contribute finite number, 
$2^{\frac{d}{2}-1} \times 2^{\frac{d}{2}-1}$, degrees of freedom to 
the corresponding creation operators, while the basis of ordinary space contribute
continuously infinite degrees of freedom. 

Is the way proposed by the {\it spin-charge-family} theory the right way to the 
next step beyond the {\it standard model}? 
The theory certainly offers a different view of the properties of fermion and boson 
fields and a different view of the second quantization of both fields than that 
offered by group theory and the second quantization by postulates.

% Sect.~\ref{fermionandgravitySCFT}.

% Correct what repeats twice!!!!
% Holger to Norma:\\ Almost compactified space, Norma
%The reduction of the number of visible dimensions is a delicate point.
% I think you have in mind the compactification type of reduction of
% dimension, but technically it looks much more like a kind of
% restriction to a brane reduction of the dimension, because you never care
% except for the gauge fields comming from Kaluza Klein for the degrees of
% freedom of a particle by moving in the extra dimension directions. For the
% fermion you take it as if there was only one relevant state or wave
% function w.r.t. the higher dimensions. The fermion could as well
% be locked on to some 3+1 dimensional brane. It can only make use of its
% spin degrees of freedom in the practical life for us.\\

% See comments of Holger from 06.01. 2021 at  17: 00

% The rest of dimensions should be nonobservable at low energies.\\

%The polynomials of the $\gamma'^a$'s form an algebra with the matrix
%multiplication as the product, denoted in this article as $*_A$.

It has happened so many times in the history of science that the simpler  
model has shown up as a more ''powerful'' one.

%
%\noindent
My working hypotheses is that the laws of nature are simple and correspondingly 
elegant and that the many body systems around the phase transitions look to us 
complicated at least from the point of view of the elementary constituents of 
fermion and boson fields.

To this working hypotheses belong also the description of the internal space
of fermions and bosons with the Clifford algebras and the simple starting action
for the (second quantized) massless fermions interacting with the (second 
quantized)~%
\footnote{Since the single fermion states, described by the Clifford odd ''basis 
vectors'',  anticommute due to the anticommuting properties of the Clifford odd 
''basis vectors'' and the single boson states, described by the Clifford even ''basis 
vectors'',  correspondingly commute there are only the second quantized fermion 
and boson fields.} 
massless bosons, representing gravity only --- the vielbeins and the two kinds 
of the spin connection fields, the gauge fields of the two kinds of the generators 
of the Lorentz transformations $S^{ab}(=\frac{i}{2}(\gamma^a \gamma^b - 
\gamma^b \gamma^a))$ and $\tilde{S}^{ab}(=\frac{i}{2}(\tilde{\gamma}^a 
\tilde{\gamma}^b - \tilde{\gamma}^b \tilde{\gamma}^a))$.

In Sect.~\ref{cliffordalgebra} I shall very shortly  overview the Clifford algebra 
description of the internal space of fermions, following Ref.~\cite{nh2021RPPNP},
and bosons (explained in my additional contribution to this 
Proceedings~\cite{norma2021SQFB}),
after the reduction of the two independent groups of Clifford algebras to only one.

In Sect.~\ref{creation} the definition of the creation and annihilation 
operators as tensor products of the ''basis vectors'' defined by the Clifford algebra
objects and basis in ordinary space is presented. 

In Sect.~\ref{achievements} the simple starting action of the {\it spin-charge-family}
theory is presented and the achievements of the theory so far discussed.

In Sect.~\ref{conclusions} the open problems of the {\it spin-charge-family}
theory are presented, and the invitation to the reader to participate.

%* Missing the paragraph as an introduction to the Clifford algebra.

% *TU 28.
%  Motivation for using Clifford algebra objects to  describe the internal
%  space of fermions with the Clifford algebra 
%  (the spin-charge-family theory and its achievements so far)\\% \label{motivation}

% *\\
% *TU 12.04. at 16:30\\
% SCFT answer: i., ii., iv., v., vi., vii., viii.,ix.,x., yes, all are spinors in $d=(13+1)$
% interacting with gravity only, iii. Clifford odd algebra,
% xi. and xii. still need to be treated.\\

% Bosons in this  Proceedings 

%  Let us start with the motivation for the review article. 
% Are the laws of nature simple and elegant manifesting complexity 
% of the many body problem
%
\section{Clifford algebra and internal space of fermions and bosons}
\label{cliffordalgebra}

I follow here Ref.~\cite{nh2021RPPNP}, Sect.~3 and also my second contribution to this 
Proceedings~\cite{norma2021SQFB}, Sect.~2.

Single fermion states are functions of  external coordinates %or momentum 
and of internal space of fermions. 
If $M^{ab}$  denote infinitesimal generators of the Lorentz algebra in both spaces,
 $M^{ab}= L^{ab} + S^{ab}$, with $L^{ab}= x^a p^b - x^b p^a$, $p^a=
i \frac{\partial}{\partial x_a}$, determining operators in ordinary space, while 
 $S^{ab}$ are equivalent operators in internal space of fermions, it follows
\begin{eqnarray}
\label{Lorentzcom}
\{M^{ab}, M^{cd}\}_{-} &=&  
 i \{M^{ad} \eta^{bc} + M^{bc} \eta^{ad} -  M^{ac} \eta^{bd}- \bf M^{bd} \eta^{ac}\}\,,\nonumber\\
\{M^{ab}, p^{c}\}_{-} &=&  -i  \eta^{ac}p^b + i \eta^{cb}p^a\,, \nonumber\\
\{M^{ab}, S^{cd}\}_{-} &=&i \{S^{ad} \eta^{bc} + S^{ad} -  S^{ac} \eta^{bd}- \bf S^{bd} \eta^{ac}\}\,,
% M^{ab \dagger} &=& \eta^{aa} \eta^{bb} {\cal {\bf M}}^{ab}\,.
\end{eqnarray}
while the Cartan subalgebra operators of the Lorentz algebra are chosen as
\begin{eqnarray}
M^{03}, M^{12}, M^{56}, \dots, M^{d-1\, d}\,, 
\label{cartanM}
\end{eqnarray}
and will be used to define the basis  in both spaces as eigenvectors of the Cartan subalgebra
members.
The metric tensor $\eta^{ab}=diag(1,-1,-1,\dots,-1,-1) $ for 
$a=(0,1,2,3,5,\dots,d)$ is used.

There are two kinds of anticommuting algebras, the Grassmann algebra  $\theta^{a}$'s 
and $p^{\theta a}$'s ($= \frac{\partial}{\partial \theta_a}$'s), in $d$-dimensional 
space  with $d$ anticommuting operators $\theta^{a}$'s and with $d$ anticommuting 
derivatives $\frac{\partial}{\partial \theta_a}$'s,
\begin{eqnarray}
\label{thetaderanti0}
\{\theta^{a}, \theta^{b}\}_{+}=0\,, \, && \,
\{\frac{\partial}{\partial \theta_{a}}, \frac{\partial}{\partial \theta_{b}}\}_{+} =0\,,
\nonumber\\
\{\theta_{a},\frac{\partial}{\partial \theta_{b}}\}_{+} &=&\delta_{ab}\,, (a,b)=(0,1,2,3,5,\cdots,d)\,,\nonumber\\
(\theta^{a})^{\dagger} &=& \eta^{a a} \frac{\partial}{\partial \theta_{a}}\,,\quad
(\frac{\partial}{\partial \theta_{a}})^{\dagger}= \eta^{a a} \theta^{a}\,,
\end{eqnarray}
where the last line was our choice~\cite{nh2018}, 
and  the two anticommuting kinds of the Clifford algebras $\gamma^a$'s  and 
$\tilde{\gamma}^{a}$'s~%
\footnote{ 
The existence of the two kinds of the Clifford algebras is discussed in~\cite{norma92,norma93,norma94,nh02,nh03}. } are expressible with the 
Grassmann algebra operators and opposite
\begin{eqnarray}
\label{clifftheta}
\gamma^{a} &=& (\theta^{a} + \frac{\partial}{\partial \theta_{a}})\,, \quad 
\tilde{\gamma}^{a} =i \,(\theta^{a} - \frac{\partial}{\partial \theta_{a}})\,,\nonumber\\
\theta^{a} &=&\frac{1}{2} \,(\gamma^{a} - i \tilde{\gamma}^{a})\,, \quad 
\frac{\partial}{\partial \theta_{a}}= \frac{1}{2} \,(\gamma^{a} + i \tilde{\gamma}^{a})\,,
%\nonumber\\
%(\gamma^{a})^{\dagger} &=&\eta^{aa}\, \gamma^{a}\,, \quad
%(\tilde{\gamma}^{a})^{\dagger} = \eta^{aa} \,\tilde{\gamma}^{a}\,,
\end{eqnarray}
offering together  $2\cdot 2^d$  operators: $2^d$ of those which are products of 
$\gamma^{a}$  and  $2^d$ of those which are products of $\tilde{\gamma}^{a}$,
the same number of operators as  of the Grassmann algebra operators. The two kinds of the
Clifford algebras anticommute, fulfilling the anticommutation relations 
\begin{eqnarray}
\label{gammatildeantiher0}
\{\gamma^{a}, \gamma^{b}\}_{+}&=&2 \eta^{a b}= \{\tilde{\gamma}^{a}, 
\tilde{\gamma}^{b}\}_{+}\,, \nonumber\\
\{\gamma^{a}, \tilde{\gamma}^{b}\}_{+}&=&0\,,\quad
 (a,b)=(0,1,2,3,5,\cdots,d)\,, \nonumber\\
(\gamma^{a})^{\dagger} &=& \eta^{aa}\, \gamma^{a}\, , \quad 
(\tilde{\gamma}^{a})^{\dagger} =  \eta^{a a}\, \tilde{\gamma}^{a}\,,\nonumber\\
% S^{ab} &=& \frac{i}{2}\, (\gamma^a \gamma^b - \gamma^b \gamma^a)\,, \quad 
% \tilde{S}^{ab} = \frac{i}{2}\, (\tilde{\gamma}^a \tilde{\gamma}^b - \tilde{\gamma}^b
% \tilde{\gamma}^a)\,,
\gamma^a \gamma^a &=& \eta^{aa}\,, \quad 
\gamma^a (\gamma^a)^{\dagger} =I\,,\quad
 \tilde{\gamma}^a  \tilde{\gamma}^a = \eta^{aa} \,,\quad
 \tilde{\gamma}^a  (\tilde{\gamma}^a)^{\dagger} =I\,,
\end{eqnarray}
where $I$ represents the unit operator. The two kinds of the Clifford algebra objects are 
obviously independent.

The corresponding infinitesimal Lorentz generators are then ${\cal {\bf S}}^{ab}$
for the Grassmann algebra, and $S^{ab}$ and $\tilde{S}^{ab}$ for the two kinds 
of the Clifford algebras.  
%It follows  for the generators of the Lorentz algebra of each of the two kinds of the Clifford 
%algebra operators, $S^{ab}$ and $\tilde{S}^{ab}$, that:
%
\begin{eqnarray}
 S^{ab} = \frac{i}{4}(\gamma^a \gamma^b - \gamma^b \gamma^a)\,,\quad  
\tilde{S}^{ab} =\frac{i}{4}(\tilde{\gamma}^a \tilde{\gamma}^b - \tilde{\gamma}^b
 \tilde{\gamma}^a)\, ,\quad &&{\cal {\bf S}}^{ab}= i \, (\theta^{a} \frac{\partial}{\partial \theta_{b}} - \theta^{b}
\frac{\partial}{\partial \theta_{a}})\,,  \nonumber\\
\{S^{ab}, \tilde{S}^{ab}\}_{-}=0\,,\quad && 
{\cal {\bf S}}^{ab} =S^{ab} + \tilde{S}^{ab}\,, \nonumber\\
\{{\cal {\bf S}}^{ab}, \, \theta^e\}_{-} = - i \,(\eta^{ae} \,\theta^b - \eta^{be}\,\theta^a)\,, \quad 
&&\{{\cal {\bf S}}^{ab}, \, p^{\theta e}\}_{-} = - i \,(\eta^{ae} \,p^{\theta b} - 
\eta^{be}\,p^{\theta a})\,,
\nonumber\\
\{ S^{ab}, \gamma^c\}_{-}&=& i (\eta^{bc} \gamma^a - \eta^{ac} \gamma^b)\,, \nonumber\\%\quad
\{ \tilde{S}^{ab}, \tilde{\gamma}^c \}_{-}&=& i (\eta^{bc} \tilde{\gamma}^a - 
\eta^{ac }\tilde{\gamma}^b)\,,\nonumber\\
\{ S^{ab}, \tilde{\gamma}^c\}_{-}&=&0\,,\quad \{\tilde{S}^{ab}, \gamma^c\}_{-}=0\,.
\label{sabtildesab}
\end{eqnarray}
The reader can find a more detailed information in Ref.~\cite{nh2021RPPNP} in Sect.~3.

% ***
It is  useful to choose the ''basis vectors'' in each of the two spaces to be products of 
eigenstates of the Cartan subalgebra members, Eq.~(\ref{cartanM}), of the Lorentz 
algebras, ($S^{ab} =\frac{i}{4}(\gamma^a \gamma^b - \gamma^b \gamma^a), 
\tilde{S}^{ab} =\frac{i}{4}(\tilde{\gamma}^a \tilde{\gamma}^b -
 \tilde{\gamma}^b  \tilde{\gamma}^a)$).
The ''eigenstates'' of each of the Cartan subalgebra members, 
Eqs.~(\ref{clifftheta}, \ref{gammatildeantiher0}), for each of the two kinds of the Clifford 
algebras separately can be found as follows,   
\begin{eqnarray}
S^{ab} \frac{1}{2} (\gamma^a + \frac{\eta^{aa}}{ik} \gamma^b) &=& \frac{k}{2}  \,
\frac{1}{2} (\gamma^a + \frac{\eta^{aa}}{ik} \gamma^b)\,,\quad
S^{ab} \frac{1}{2} (1 +  \frac{i}{k}  \gamma^a \gamma^b) = \frac{k}{2}  \,
 \frac{1}{2} (1 +  \frac{i}{k}  \gamma^a \gamma^b)\,,\nonumber\\
\tilde{S}^{ab} \frac{1}{2} (\tilde{\gamma}^a + \frac{\eta^{aa}}{ik} \tilde{\gamma}^b) &=& 
\frac{k}{2}  \,\frac{1}{2} (\tilde{\gamma}^a + \frac{\eta^{aa}}{ik} \tilde{\gamma}^b)\,,
\quad
\tilde{S}^{ab} \frac{1}{2} (1 +  \frac{i}{k}  \tilde{\gamma}^a \tilde{\gamma}^b) = 
 \frac{k}{2}  \, \frac{1}{2} (1 +  \frac{i}{k} \tilde{\gamma}^a \tilde{\gamma}^b)\,,
\label{eigencliffcartan}
\end{eqnarray}
 $k^2=\eta^{aa} \eta^{bb}$.
 The proof of Eq.~(\ref{eigencliffcartan}) is presented in App.~(I) of Ref.~\cite{nh2021RPPNP}, 
 Statement 2a.
The Clifford "basis vectors" --- nilpotents $\frac{1}{2} (\gamma^a + \frac{\eta^{aa}}{ik} \gamma^b), (\frac{1}{2} (\gamma^a + \frac{\eta^{aa}}{ik} \gamma^b))^2=0$ and 
projectors $ \frac{1}{2} (1 +  \frac{i}{k}  \tilde{\gamma}^a \tilde{\gamma}^b),
( \frac{1}{2} (1 +  \frac{i}{k}  \tilde{\gamma}^a \tilde{\gamma}^b))^2 =
 \frac{1}{2} (1 +  \frac{i}{k}  \tilde{\gamma}^a \tilde{\gamma}^b)$ --- 
of both algebras are normalized,  up to a phase, as described in the contribution 
of the same outhor in this Proceedings~\cite{norma2021SQFB}.

Both, nilpotents and projectors, have half integer spins. 

It is useful to introduce the notation for the "eigenvectors" of the 
two Cartan subalgebras as follows, Ref.~\cite{nh02,nh03},
\begin{eqnarray}
\stackrel{ab}{(k)}:&=& 
\frac{1}{2}(\gamma^a + \frac{\eta^{aa}}{ik} \gamma^b)\,,\quad 
\stackrel{ab}{(k)}^{\dagger} = \eta^{aa}\stackrel{ab}{(-k)}\,,\quad 
(\stackrel{ab}{(k)})^2 =0\,, \quad \stackrel{ab}{(k)}\stackrel{ab}{(-k)}
=\eta^{aa}\stackrel{ab}{[k]}\nonumber\\
\stackrel{ab}{[k]}:&=&
\frac{1}{2}(1+ \frac{i}{k} \gamma^a \gamma^b)\,,\quad \;\,
\stackrel{ab}{[k]}^{\dagger} = \,\stackrel{ab}{[k]}\,, \quad \quad \quad \quad
(\stackrel{ab}{[k]})^2 = \stackrel{ab}{[k]}\,, 
\quad \stackrel{ab}{[k]}\stackrel{ab}{[-k]}=0\,,
\nonumber\\
\stackrel{ab}{(k)}\stackrel{ab}{[k]}& =& 0\,,\qquad \qquad \qquad 
\stackrel{ab}{[k]}\stackrel{ab}{(k)}=  \stackrel{ab}{(k)}\,, \quad \quad \quad
  \stackrel{ab}{(k)}\stackrel{ab}{[-k]} =  \stackrel{ab}{(k)}\,,
\quad \, \stackrel{ab}{[k]}\stackrel{ab}{(-k)} =0\,.
% \nonumber\\
%
% \stackrel{ab}{\tilde{(k)}}:&=& 
% \frac{1}{2}(\tilde{\gamma}^a + \frac{\eta^{aa}}{ik} \tilde{\gamma}^b)\,,\quad 
% \stackrel{ab}{\tilde{(k)}}^{\dagger} = \eta^{aa}\stackrel{ab}{\tilde{(-k)}}\,,\quad
% (\stackrel{ab}{\tilde{(k)}})^2=0\,,\nonumber\\
% \stackrel{ab}{\tilde{[k]}}:&=&
% \frac{1}{2}(1+ \frac{i}{k} \tilde{\gamma}^a \tilde{\gamma}^b)\,,\quad \;\,
% \stackrel{ab}{\tilde{[k]}}^{\dagger} = \,\stackrel{ab}{\tilde{[k]}}\,,
% \quad \quad \quad \quad
% (\stackrel{ab}{\tilde{[k]}})^2=\stackrel{ab}{\tilde{[k]}}\,,\nonumber\\
%
% \stackrel{ab}{\tilde{(k)}}\stackrel{ab}{\tilde{[k]}}& =& 0\,,\qquad \qquad \qquad 
% \stackrel{ab}{\tilde{[k]}}\stackrel{ab}{\tilde{(k)}}=  \stackrel{ab}{\tilde{(k)}}\,, 
% \quad \quad \quad
%   \stackrel{ab}{\tilde{(k)}}\stackrel{ab}{\tilde{[-k]}} =  \stackrel{ab}{\tilde{(k)}}\,,
% \quad \, \stackrel{ab}{\tilde{[k]}}\stackrel{ab}{\tilde{(-k)}} =0\,,
% \nonumber\\
\label{graficcliff}
\end{eqnarray}
The corresponding expressions for nilpotents $\stackrel{ab}{\tilde{(k)}}$ and projectors 
$\stackrel{ab}{\tilde{[k]}}$ follows if we replace in Eq.~(\ref{graficcliff})  $\gamma^a$'s
by  $\tilde{\gamma}^a$'s,  the same relation $k^2 = \eta^{aa} \eta^{bb}$ 
is valid for both algebras. 

Let us notice that the ''eigenvectors'' of  the Cartan subalgebras  are equivalent and the eigenvalues are the same in both algebras: Both algebras have projectors and nilpotents:
($ (\stackrel{ab}{[k]})^2= \stackrel{ab}{[k]}\,,(\stackrel{ab}{(k)})^{2}=0$),
($ (\stackrel{ab}{\tilde{[k]}})^2= \stackrel{ab}{\tilde{[k]}}\,,
 (\stackrel{ab}{\tilde{(k)}})^{2}=0$).

In each of the two independent algebras we have  two groups of $2^{\frac{d}{2}-1}$ 
members which are eigenvectors of all the Cartan  subalgebra members, Eq.~(\ref{cartanM}), 
appearing in $2^{\frac{d}{2}-1}$ irreducible representations which have an odd Clifford
character --- they are products of an odd number of $\gamma^a$'s ($\tilde{\gamma}^a$'s). 
These two groups are Hermitian conjugated to each other.  We make a choice of one 
of the two groups of the Clifford odd ''basis vectors'' and name these   ''basis vectors''
$\hat{b}^{m \dagger}_{f}$, $m$ describing $2^{\frac{d}{2}-1}$ members of one
 irreducible representation, $f$ describing one of $2^{\frac{d}{2}-1}$  irreducible 
 representations. The $2^{\frac{d}{2}-1}\times $  $2^{\frac{d}{2}-1}$ members 
 of the second group, Hermitian conjugated to  $\hat{b}^{m \dagger}_{f}$, are 
 named as $\hat{b}^{m}_{f}= $  $(\hat{b}^{m \dagger}_{f})^{\dagger}$. 
 
There are besides two Clifford odd groups in each of the two algebras  $\gamma^a$'s 
and $\tilde{\gamma}^a$'s, also two Clifford even groups.  They are superposition of an 
even number of $\gamma^a$'s ($\tilde{\gamma}^a$'s).  I named these  two 
$2^{\frac{d}{2}-1}\times $  $2^{\frac{d}{2}-1}$ Clifford  even ''basis vectors''
${\hat{\cal A}}^{m \dagger}_{f}$   and ${\hat{\cal B}}^{m \dagger}_{f}$, respectively.
${\hat{\cal A}}^{m \dagger}_{f}$ represent gauge vectors of $\hat{b}^{m \dagger}_{f}$,
on which they operate. ${\hat{\cal B}}^{m \dagger}_{f}$ operate on $\hat{b}^{m}_{f}$.
I discuss their properties  in my second contribution of this 
Proceedings~\cite{norma2021SQFB}.  

%*** 13.11.2021 at 14:00

The ''basis vectors'' of an odd Clifford character,  %Clifford odd ''basis vectors'' 
$\hat{b}^{m \dagger}_{f}$, and their Hermitian conjugated partners, $\hat{b}^{m }_{f}$,
fulfil the postulates for second quantized fermions of Dirac, if we reduce both Clifford algebras to only one~\cite{PRD2018,n2019PIPII,IARD2020}, while keeping all the relations, presented in Eq.~(\ref{gammatildeantiher0}), valid. Let us make a choice of $\gamma^a$'s and
postulate the application of  $\tilde{\gamma}^a$'s on $B$ which is a superposition of any 
products of $\gamma^a$'s as follows
\begin{eqnarray}
\{\tilde{\gamma}^a B &=&(-)^B\, i \, B \gamma^a\}\, |\psi_{oc}>\,,
\label{tildegammareduced}
\end{eqnarray}
with $(-)^B = -1$, if $B$ is (a function of) an odd products of $\gamma^a$'s,
 otherwise $(-)^B = 1$~\cite{nh03}, $|\psi_{oc}>$ is defined in 
Eq.~(\ref{vaccliff}). (Sects.~(2.1, 2.2 in~\cite{norma2021SQFB}) and Sects.~(3.2.2, 3.2.3
in~\cite{nh2021RPPNP})). 

The vacuum state $ |\psi_{oc}>$ is defined as follows
\begin{eqnarray}
\label{vaccliff}
|\psi_{oc}>= \sum_{f=1}^{2^{\frac{d}{2}-1}}\,\hat{b}^{m}_{f}{}_{*_A}
\hat{b}^{m \dagger}_{f} \,|\,1\,>\,,
\end{eqnarray}
for one of the members $m$, anyone, of the odd irreducible representation $f$,
with $|\,1\,>$, which is the vacuum without any structure, the identity, 
$\, {}_{*_A}$ means the algebraic product.
It follows that $\hat{b}^{m}_{f}{}_{*_A} |\psi_{oc}>=0$ and 
$\hat{b}^{m \dagger}_{f}{}_{*_A}\,|\psi_{oc}>= |\psi^{m}_{f}>$.

 After the postulate of Eq.~(\ref{tildegammareduced}) ''basis vectors'' 
 $\hat{b}^{m \dagger}_{f}$ which are  
superposition of an odd products of $\gamma^a$'s  (represented by an odd number of 
nilpotents, the  rest are projectors) obey all the fermions second 
quantization postulates of Dirac. There are $\tilde{S}^{ab}$ which dress the 
irreducible representations with the family quantum numbers of the Cartan subalgebra
members  ($\tilde{S}^{03}, \tilde{S}^{12}, \tilde{S}^{56}, \dots, \tilde{S}^{d-1\,d}$), Eq.~(\ref{cartanM}). 
%%%%%%%%%%

% Let us write down the anticommutation relations of Clifford odd "basic vectors",
% representing the creation operators in internal space of fermions with half integer
% spin represented by $\gamma^a$'s and of the corresponding Hermitian
% conjugated partners annihilation operators again. After the reduction of the 
% Clifford algebra any irreducible representation carry the family quantum number, 
% distinguishing families among themselves.
%
\begin{eqnarray}
\{ \hat{b}^{m}_{f}, \hat{b}^{m' \dagger}_{f'} \}_{*_{A}+}\, |\psi_{oc}> 
&=& \delta^{m m'} \, \delta_{ff'} \,  |\psi_{oc}>\,,\nonumber\\
\{ \hat{b}^{m}_{f}, \hat{b}^{m'}_{f'} \}_{*_{A}+}  \,  |\psi_{oc}>
&=& 0 \,\cdot\,  |\psi_{oc}>\,,\nonumber\\
\{\hat{b}^{m  \dagger}_{f},\hat{b}^{m' \dagger}_{f'}\}_{*_{A}+} \, |\psi_{oc}>
&=& 0 \, \cdot\, |\psi_{oc}>\,,\nonumber\\
 \hat{b}^{m \dagger}_{f} \,{}_{*_{A}} |\psi_{oc}>&=& |\psi^{m}_{f}>\,, \nonumber\\
 \hat{b}^{m}_{f}   \,{*_{A}}  |\psi_{oc}>&=& 0 \,\cdot\,  |\psi_{oc}>\,,
\label{alphagammatildeprod}
\end{eqnarray}
with ($m,m'$) denoting the "family" members and ($f,f'$) denoting "families",
${*_{A}}$ represents the algebraic multiplication of $ \hat{b}^{\dagger m}_{f} $
and  $ \hat{b}^{m}_{f} $ with the vacuum state $|\psi_{oc}>$ of 
Eq.~(\ref{vaccliff}) and among themselves, taking into account 
Eq.~(\ref{gammatildeantiher0}).
%

%%%%%%%%%%

 Ref.~(\cite{norma2021SQFB}, Sects.~2.4 and 3) presents the starting study of properties 
 of the second quantized boson fields, the internal space of which is represented by the 
''basis vectors'' ${\hat{\cal A}}^{m \dagger}_{f}$ which appear as the gauge fields of 
the second quantized fermion fields the internal space of which is described by the ''basis 
vectors'' $\hat{b}^{m \dagger}_{f}$.

% *** Sections 2.4 mainly Sect. 3. shortly of \cite{norma2021SQFB}

% ***\\
% *** 05.11.2021 at 20:20\\
 
We pay attention on even dimensional spaces, $d=2(2n+1)$ or $d=4n$, $n\ge0$, only.

\section{Creation and annihilation operators}
\label{creation}

Here Sect.~3.3 of Ref.~\cite{nh2021RPPNP} is roughly followed.

Describing fermion fields as the creation ${\bf \hat{b}}^{s \dagger}_{f} (\vec{p})$ 
and annihilation ${\bf \hat{b}}^{s}_{f} (\vec{p})$ operators operating on the
vacuum state we make  tensor products, ${}_{*_T}$, of $2^{\frac{d}{2}-1}\times $
$2^{\frac{d}{2}-1}$ Clifford odd ''basis vectors'' $\hat{b}^{m \dagger}_{f}$  and of 
continuously infinite basis in ordinary space determined by $\hat{b}^{\dagger}_{\vec{p}}$
%        \label{wholespacegeneral}
 \begin{eqnarray}
\label{wholespacegeneral}
\{{\bf \hat{b}}^{s \dagger}_{f} (\vec{p}) \,&=& \sum_{m}\, c^{m s}{}_{f}(\vec{p})\,
\hat{b}^{\dagger}_{\vec{p}}\,{}_{*_{T}}\,\hat{b}^{m \dagger}_{f}\} \,
|\psi_{oc}>\,*_{T}\, |0_{\vec{p}}> \,,                                                                                                 
 \end{eqnarray}
where $\vec{p}$ determines the momentum in ordinary space with $p^0 =|\vec{p}|$ 
and $s$ determines all the rest of quantum numbers. 
The state $|\psi_{oc} >\,*_{T}\, |0_{\vec{p}} >$ is considered as the vacuum for a 
starting  single particle state from which one obtains the other 
single particle state by the operators,  $\hat{b}_{\vec{p}}$, which pushes the 
momentum by an amount $\vec{p}$, in a tensor product with  $\hat{b}^{m \dagger}_{f}$.
We have
\begin{eqnarray}
\label{creatorp}
|\vec{p}>&=& \hat{b}^{\dagger}_{\vec{p}} \,|\,0_{p}\,>\,,\quad 
<\vec{p}\,| = <\,0_{p}\,|\,\hat{b}_{\vec{p}}\,, \nonumber\\
<\vec{p}\,|\,\vec{p}'>&=&\delta(\vec{p}-\vec{p}')=
<\,0_{p}\,|\hat{b}_{\vec{p}}\; \hat{b}^{\dagger}_{\vec{p}'} |\,0_{p}\,>\,, 
\nonumber\\
&&{\rm leading \;to\;} \nonumber\\
\hat{b}_{\vec{p'}}\, \hat{b}^{\dagger}_{\vec{p}} &=&\delta(\vec{p'}-\vec{p})\,,
\end{eqnarray}
since we normalize $<\,0_{p}\, |\,0_{p}\,>=1$ to identity.

The ''basis vectors'' $\hat{b}^{m \dagger}_{f}$  which are products of an odd number of nilpotent, the rest to $\frac{d}{2}$  are then projectors, anticommute, transferring the anticommutativity to  the creation operators ${\bf \hat{b}}^{s \dagger}_{f} (\vec{p}) $ 
and correspondingly also to their Hermitian conjugated partners annihilation operators  
${\bf \hat{b}}^{s}_{f} (\vec{p})$, Eq.~(\ref{wholespacegeneral}).  The creation
and annihilation operators then fulfil the anticommutation relations of the second quantized
fermions explaining the postulates of Dirac
\begin{eqnarray}
\{  \hat{\bf b}^{s' }_{f `}(\vec{p'})\,,\, 
\hat{\bf b}^{s \dagger}_{f }(\vec{p}) \}_{+} \,|\psi_{oc}> |0_{\vec{p}}>&=&
\delta^{s s'} \delta_{f f'}\,\delta(\vec{p}' - \vec{p})\, |\psi_{oc}> |0_{\vec{p}}>
\,,\nonumber\\
\{  \hat{\bf b}^{s' }_{f `}(\vec{p'})\,,\, 
\hat{\bf b}^{s}_{f }(\vec{p}) \}_{+} \,|\psi_{oc}> |0_{\vec{p}}>&=&0\,
 |\psi_{oc}> |0_{\vec{p}}>
\,,\nonumber\\
\{  \hat{\bf b}^{s' \dagger}_{f '}(\vec{p'})\,,\, 
\hat{\bf b}^{s \dagger}_{f }(\vec{p}) \}_{+}\, |\psi_{oc}> |0_{\vec{p}}>&=&0
\,|\psi_{oc}> |0_{\vec{p}}>
\,,\nonumber\\
 \hat{\bf b}^{s \dagger}_{f }(\vec{p}) \,|\psi_{oc}> |0_{\vec{p}}>&=&
|\psi^{s}_{f}(\vec{p})>\,\nonumber\\
 \hat{\bf b}^{s}_{f }(\vec{p}) \, |\psi_{oc}> |0_{\vec{p}}>&=&0
 \,|\psi_{oc}> |0_{\vec{p}}>\nonumber\\
 |p^0| &=&|\vec{p}|\,.
\label{Weylpp'comrel}
\end{eqnarray}
%

%

% To  describe the 
% ''basis vectors'' of the corresponding boson fields we choose even number of nilpotents,
% %the number of both is again $\frac{d}{2}$. Correspondingly odd ''basis vectors 
% anticommute and  even ''basis vectors'' commute.

{\bf Statement } {\it The description of the internal space of fermions with the 
superposition of odd products of $\gamma^a$'s, that is with the clifford odd 
''basis vectors'', not only explains the Dirac's postulates of the second quantized fermions
but also  explains the appearance of families of fermions.}

Ref.~\cite{norma2021SQFB} is offering the explanation for the second quantized 
commuting boson fields (described by the ''basis vectors'' of an even number of 
nilpotents, the rest are projectors), they are the gauge fields of the anticommuting 
fermion fields (described by the ''basis vectors'' of an odd number of nilpotents).

%*** 13.11.2021 at 20:41

%
\section{Achievements so far of {\it spin-charge-family} theory}
\label{achievements}

Here Sects.~(6, 7.2.2 and 7.3.1) of Ref.~\cite{nh2021RPPNP}, which review 
shortly the achievements so far of the  {\it spin-charge-family} theory, are  
followed. 

The main new achievement of this theory in the last few years is the recognition 
that the description of the internal space of fermion fields with the Clifford algebra 
objects in $d > (3+1)$ not only offers the explanation for all the assumptions of the 
{\it standard model} for fermion and boson fields, with the appearance of families 
for fermion fields and the properties of the corresponding vector and scalar gauge 
fields included, but also get to know, 
that the anticommuting property of the internal space of fermions takes care
of the second quantization properties of fermions, so that the second quantized 
postulates are not needed. The second  quantized properties of fermions origin
in their internal space and are transferred to creation and annihilation operators.
This year contribution to Proceedings~Ref.~\cite{norma2021SQFB} offers the
recognition that also commuting properties of the second quantized boson fields 
origin in the internal space of bosons. 

Describing the internal space of bosons by the Clifford even ''basis vectors'', 
written in terms of the Clifford even number of $\gamma^a$'s, these Clifford even 
''basis vectors'', ${\hat{\cal A}}^{m \dagger}_{f}$, applying on fermion states 
transform the ''basis vectors''  $\hat{b}^{m \dagger}_f$ either into another 
''basis vectors'' $\hat{b}^{m' \dagger}_f$ with the same family quantum number 
$f$, or if written in terms of  the Clifford even number of $\tilde{\gamma}^a$'s,
${\hat{\cal \tilde{A}}}^{m \dagger}_{f}$, transform $\hat{b}^{m' \dagger}_f$ 
to $\hat{b}^{m \dagger}_{f `}$, keeping the family member quantum number 
$m$ unchanged and changing the family quantum number to $f `$.~\footnote{ 
The first operation happens if the internal space of bosons is described by ''basis
vectors'' which are even products of nilpotents of the kind 
${\hat{\cal A}}^{m \dagger}_{f} = \stackrel{03}{(-i)} \stackrel{12}{(-)} 
\stackrel{56}{[+]}\cdots  \stackrel{d-1\, d}{[+]})$, in this particular case  
two nilpotents form ''basis vectors'', the 
second operation happens if  all the nilpotents $ \stackrel{a b}{(k)}$ and projectors 
$ \stackrel{cd}{[k]}$ are replaced by the corresponding $ \stackrel{a b}{\tilde{(k)}}$
and  $ \stackrel{a b}{\tilde{[k]}}$, respectively.} This topic, started in 
 Ref.~\cite{norma2021SQFB}, needs further study. 

The {\it spin-charge-family} theory proposes a simple action for interacting second quantized massless fermions and the corresponding gauge fields in $d=(13+1)$-dimensional space as 
\begin{eqnarray}
{\cal A}\,  &=& \int \; d^dx \; E\;\frac{1}{2}\, (\bar{\psi} \, \gamma^a p_{0a} \psi) 
+ h.c. +
%{\mathcal L}_{f} +  
\nonumber\\  
               & & \int \; d^dx \; E\; (\alpha \,R + \tilde{\alpha} \, \tilde{R})\,,
\nonumber\\
               p_{0a } &=& f^{\alpha}{}_a p_{0\alpha} + \frac{1}{2E}\, \{ p_{\alpha},
E f^{\alpha}{}_a\}_- \,,\nonumber\\
          p_{0\alpha} &=&  p_{\alpha}  - \frac{1}{2}  S^{ab} \omega_{ab \alpha} - 
                    \frac{1}{2}  \tilde{S}^{ab}   \tilde{\omega}_{ab \alpha} \,,
                    \nonumber\\                    
R &=&  \frac{1}{2} \, \{ f^{\alpha [ a} f^{\beta b ]} \;(\omega_{a b \alpha, \beta} 
- \omega_{c a \alpha}\,\omega^{c}{}_{b \beta}) \} + h.c. \,, \nonumber \\
\tilde{R}  &=&  \frac{1}{2} \, \{ f^{\alpha [ a} f^{\beta b ]} 
\;(\tilde{\omega}_{a b \alpha,\beta} - \tilde{\omega}_{c a \alpha} \,
\tilde{\omega}^{c}{}_{b \beta})\} + h.c.\,.               
\label{wholeaction}
\end{eqnarray}
Here~\footnote{$f^{\alpha}{}_{a}$ are inverted vielbeins to 
$e^{a}{}_{\alpha}$ with the properties $e^a{}_{\alpha} f^{\alpha}{\!}_b = 
\delta^a{\!}_b,\; e^a{\!}_{\alpha} f^{\beta}{\!}_a = \delta^{\beta}_{\alpha} $, 
$ E = \det(e^a{\!}_{\alpha}) $.
Latin indices  
$a,b,..,m,n,..,s,t,..$ denote a tangent space (a flat index),
while Greek indices $\alpha, \beta,..,\mu, \nu,.. \sigma,\tau, ..$ denote an Einstein 
index (a curved index). Letters  from the beginning of both the alphabets
indicate a general index ($a,b,c,..$   and $\alpha, \beta, \gamma,.. $ ), 
from the middle of both the alphabets   
the observed dimensions $0,1,2,3$ ($m,n,..$ and $\mu,\nu,..$), indexes from 
the bottom of the alphabets
indicate the compactified dimensions ($s,t,..$ and $\sigma,\tau,..$). 
We assume the signature $\eta^{ab} =
diag\{1,-1,-1,\cdots,-1\}$.} 
$f^{\alpha [a} f^{\beta b]}= f^{\alpha a} f^{\beta b} - f^{\alpha b} f^{\beta a}$.

This simple action in $d=(13+1)$-dimensional space, \\
{\bf i.} in which massless fermions 
interact with the massless gravitation fields only (with the vielbeins and the two kinds 
of the spin connection fields, the gauge fields of $S^{ab}$ and $\tilde{S}^{ab}$, 
respectively),\\
{\bf ii.} together with the assumption that the internal space of the second 
quantized fermions are described by the Clifford odd ''basis vectors'' (what explains after 
the break of symmetries at low energies the appearance of quarks and leptons and 
antiquarks and antileptons of the {\it standard model} and  the existence 
 of families, predicting the number of families~\cite{gn2014}), \\ 
 %after breaking symmetries at low energy regime), 
{\bf iii.}and the internal space of the second quantized boson fields are described by the 
Clifford even ''basis vectors'', offers the explanations for \\
{\bf iv.} not only all the assumptions of 
the {\it standard model} --- for properties of quarks and leptons and antiquarks and 
antileptons (explaining the relations among spins, handedness and charges of fermions 
and antifermions~\cite{norma2001,IARD2016}) and for 
the appearance of families of quarks and leptons~\cite{nh02,nh03,pikan2006}, \\
{\bf v.}  for the second quantized postulates of Dirac~\cite{prd2018,n2019PIPII}, \\
{\bf vi.} for the appearance of the vector gauge fields to the corresponding  fermion 
fields~\cite{nd2017}, \\
{\bf vii.} for the appearance of gauge scalars explaining the interactions among fermions belonging to different 
families~\cite{n2012scalars,JMP2013,normaJMP2015,mdn2006,gmdn2007,gmdn2008,%
gn2014,nd2017}, and correspondingly of the appearance of 
the higgs scalar and Yukawa couplings, \\
{\bf viii.} predicting the number of families --- the fourth one to the observed 
three~\cite{gn2014},\\
{\bf ix.}  predicting the second group of four families the stable of which explains the appearance of the {\it dark matter}~\cite{gn2009,IARD2016}, \\
{\bf x.} predicting additional gauge fields, \\
{\bf xi.} predicting additional scalar fields, which explain the existence of matter/antimatter
asymmetry~\cite{n2014matterantimatter},\\
 and several others.

The manifold $M^{(13 +1)}$ breaks at high scale $\propto 10^{16}$ GeV or higher
first to $M^{(7 +1)}\times M^{(6)}$ due to the appearance of the scalar condensate 
(so far just assumed, not yet proven that it appears spontaneously)  of the two right 
handed neutrinos with the family quantum numbers of the group of four families, 
which does not include the observed three families 
%(Table~\ref{Table III.}),Sect.~\ref{vectorscalar3+1}, 
bringing masses (of the scale of break $\propto 10^{16}$ GeV or higher) to all the 
gauge fields,  which interact with the condensate~\cite{n2014matterantimatter}.
%Sect.~\ref{actionGrassCliff}. 

% It is shown in Ref.~\cite{nd2017,IARD2020} that the spin connection gauge fields 
% manifest in $d=(3+1)$ as the ordinary gravity, the known vector gauge fields and 
% the scalar gauge fields, offering the (simple) explanation for the origin of higgs  
% assumed by the {\it standard model}, explaining as well the Yukawa 
% couplings. 

% The theory predicts new vector and scalar gauge fields, 
% Sect.~\ref{vectorscalar3+1}, what offers explanation for the 
% {\it dark matter}~\cite{gn2009,IARD2016}, Sect.~\ref{predictionSCFT} and 
% for the {\it matter-antimatter asymmetry}~\cite{n2014matterantimatter} in the 
% universe, Sect.~\ref{scalar3+1}.

%***22.02.2021 at 19:30\\
%%%*Shorten and make easier to read this part, up and down.

% ***\\
% *** 06.2021 at 14:31\\

Since the left handed spinors --- fermions --- couple differently (with respect to 
$M^{(7+1)}$) to scalar fields than the right handed ones, the break can leave massless 
and mass protected $2^{((7+1)/2-1)}(= 8)$ families~\cite{NHD}.
%, Sect.\ref{TDN0}, Eq.~(\ref{weylTDN}). 
The rest of families get heavy masses~\footnote{%
A toy model~\cite{NHD,ND012,nh2008} was studied in 
$d=(5+1)$-dimensional space with the action presented in Eq.~(\ref{wholeaction}),
%Sect.\ref{TDN0}, Eq.~(\ref{weylTDN}). 
%
The break from $d=(5+1)$ 
to $d=(3+1) \times$ an almost $S^{2}$ was studied for a particular choice of 
vielbeins and for a class of spin connection fields. While the manifold $M^{(5+1)}$ 
breaks into $M^{(3+1)}$ times an almost $S^2$ the $2^{((3+1)/2-1)}$ families 
remain massless and mass protected. 
Equivalent assumption, although not yet proved how does it 
really work, is made also for the $d=(13+1)$ case. This study is in progress quite 
some time.}. 

 The manifold $M^{(7+1)} \times SU(3)\times U(1)$  breaks further by %the assumption following the
%{\it standard model} assumption for the electroweak break with 
the scalar fields, presented in Sect.~\ref{scalar3+1}, 
% which are the part of the starting action,
 to $M^{(3+1)} \times$  $SU(3)\times U(1)$
%, the last   $SU(2)\times SU(2)$ included in $M^{(4)}$ 
at the electroweak break. This happens since the 
scalar fields with the space index $(7,8)$,  Subsubsect.~\ref{scalar3+1}, they
 are a part of  a simple 
starting action~Eq.(\ref{wholeaction}), gain the constant values (the nonzero vacuum 
expectation values independent of the coordinates in $d=(3+1)$).
These scalar fields carry with respect to the space index the weak charge 
$\pm \frac{1}{2}$ and the hyper charge 
$\mp \frac{1}{2}$~\cite{n2014matterantimatter,IARD2016}, 
Sect.~\ref{scalar3+1}, just as required by the {\it standard model}, 
manifesting with respect to $\tilde{S}^{ab}$ and $S^{ab}$ additional quantum 
numbers.

Let us point out that all the fermion fields (with the families of fermions 
and the neutrinos
forming the condensate included), the vector and the scalar gauge fields, offering
explanation for  by the {\it standard model}  postulated ones, origin in the 
simple starting action. 

The starting action, Eq.~(\ref{wholeaction}), has only a few parameters.
It is assumed that the  coupling of fermions to $\omega^{ab}{}_{c}$'s 
can differ  from the coupling of fermions to $\tilde{\omega}^{ab}{}_{c}$'s, 
The reduction of the Clifford space, Eq.~\ref{tildegammareduced}, causes
this difference. The additional breaks of symmetries influence the coupling 
constants in addition. 

The breaks of symmetries is under consideration for quite 
a long time and  has not yet been finished.

All the observed properties of fermions, of vector gauge fields 
and scalar gauge fields follow from the simple starting action, while the 
breaks of symmetries influence the properties of fermion and boson fields
as well. 

\subsection{Properties of  interacting massless fermions as manifesting in $d=(3+1)$
before electroweak break}
\label{fermionactionSCFT}
% 6.1. before 4.3.2.

One irreducible representation of $SO(13,1)$ includes all the left handed and right handed
quarks and leptons and antiquarks and antileptons as one can see in Table~5 of Ref.~%
\cite{norma2021SQFB} in this Proceedings or in Table~7 of Ref.~\cite{nh2021RPPNP}. 
In both tables fermion ''basis vectors'' are represented by odd numbers of nilpotents 
and their properties analysed  from the point of view of the {\it standard model} 
subgroups $SO(3,1)\times SU(2)\times SU(2) \times SU(3)\times U(1)$ of the group 
$SO(13,1)$. Quarks and leptons as well as antiquarks and antileptons
 appear with handedness as required by the {\it standard model}.

One easily notices that quarks and leptons have the same content of the 
subgroup $SO(7,1)$, distinguishing only in $SU(3)\times U(1)$ content of $SO(6)$: 
all the quarks, left and right handed, have the ''fermion'' $\tau^4 $ equal to 
$\frac{1}{6}$ and appear in three colours, all the leptons, left and right handed, 
have $\tau^4 $ equal to $- \frac{1}{2}$  and are colourless. 

Also antiquarks and antileptons have the same content of the subgroup 
$SO(7,1)$ (which is different from the one of quarks and leptons), and differ in  
$SU(3)\times U(1)$ content of $SO(6)$, all the antiquarks, left and right handed, have 
$\tau^4 $ equal to $-\frac{1}{6}$ and appear in three anticolours, all the antileptons 
have $\tau^4 $ equal to $\frac{1}{2}$  and are anticolourless. 

Let us notice also that since there are two $SU(2)$ weak charges the  right handed 
neutrinos and the left handed antineutrinos have non zero  the second $SU(2)_{II}$
weak charge and interact with the $SU(2)_{II}$ weak field. Both have the 
{\it standard model} hyper charge $Y=\tau^4 + \tau^{23}$ equal to zero.
Let me point out that this particular property are offered also by the $SO(10)$ unifying 
model~\cite{FritzMin}, but with the manifold $M^(3+1)$ decoupled from
charges. (Comments can be found in Ref.~\cite{nh2021RPPNP}, Sect.~7).

The expressions for the generators of the Lorentz transformations of subgroups 
$SO(3,1)\times SU(2)\times SU(2) \times SU(3)\times U(1)$ of the group $SO(13,1)$
can be found in App.~\ref{generators} (also in Eqs.~(39-41) of Ref.~\cite{norma2021SQFB} 
or in Eqs.~(85-89) of Ref.~\cite{nh2021RPPNP}).

The condensate, presented in  Table~\ref{Table con.}  
(Table~6 of Ref.~\cite{nh2021RPPNP}), makes 
one of the two weak $SU(2)$ fields massive and causes the break of symmetries from $M^{(13+1)}$ to $M^{(7+1)}\times SU(3)\times U(1)$~\cite{NHD,ND012,nh2008}, 
leaving only two decoupled groups of  four families massless, $2^{\frac{7+1}{2}-1}=
8$. The reader can find these two groups of families in 
Table~\ref{Table III.} (from Table~5 of Ref.~\cite{nh2021RPPNP}).

% Let us present here  Table~5 and Table~6 of Ref.~\cite{nh2021RPPNP}.

Table~\ref{Table III.} presents ''basis vectors'' ($\hat{b}^{m \dagger}_{f}$, Eq.~(\ref{alphagammatildeprod})) for eight families of the right handed $u$-quark 
of  the colour  $(\frac{1}{2}, \frac{1}{2\sqrt{3}})$ and the right handed colourless
$\nu$-lepton. The SO(7,1) content of the $SO(13,1)$ group are in both cases 
identical, they 
distinguish only in the $SU(3)$ and $U(1)$ subgroups of $SO(6)$. All the members 
of any of these eight families of Table~\ref{Table III.} follows from either the $u$-quark 
or the $\nu$-lepton  by the application of $S^{ab}$. Each family carries the family quantum
numbers, determined by the Cartan subalgebra of $\tilde{S}^{ab}$ in Eq.~(\ref{cartanM}) 
and presented in Table~\ref{Table III.}. 
% When we treat the $d=(9+1)$ case, the families can be assumed, determined in 
% this case with the Cartan subalgebra members of $\tilde{\tau}^{13}, 
% \tilde{\tau}^{23}, \tilde{N}^{3}_L,\tilde{N}^{3}_R$ and $\tilde{S}^{9 \,10}$. 
% This is not the  case, which would be realized in nature, at least it is not yet observed.

The two groups of families are after the break of symmetries decoupled since 
$\{\tilde{N}^{i}_{L},\, \tilde{N}^{j}_{R}\}_{-}=0\, ,\forall (i,j)$, 
$\{\tilde{\tau}^{1\,i},\, \tilde{\tau}^{2\,j}\}_{-}=0\, ,\forall (i,j)$,
$\{\tilde{N}^{i}_{L,R},\,\tilde{\tau}^{1,2\,j}\}_{-}=0\, ,\forall (i,j)$, 
while $\{S^{ab}, \, \tilde{S}^{cd}\,\}_{-}=0$, 
since $\{\gamma^a, \,\tilde{\gamma}^a\}_{-}=0$, Eq.~(\ref{gammatildeantiher0}).

% **** 6,1,7.2.2, 7.3.1.\\

% ***  Check Tables 5 and 6\\
% *** 07.11.2021 at 12:15\\
% *** \\

%
 \begin{table}
 \begin{center}
   \begin{tiny}
\begin{minipage}[t]{16.5 cm}
%%\begin{small}
%\begin{tiny}
\caption{Eight families of the ''basis vectors'' $\hat{b}^{m \dagger}_{f}$,  
of $\hat{u}^{c1 \dagger}_{R}$ --- the right 
handed $u$-quark with spin $\frac{1}{2}$ and the colour charge $(\tau^{33}=1/2$, 
$\tau^{38}=1/(2\sqrt{3}))$, appearing in the first line of Table~7 in 
Ref.~\cite{nh2021RPPNP}, or Table~5 in Ref.~\cite{norma2021SQFB}
%\ref{Table so13+1.} 
--- and of  the colourless right handed neutrino $\hat{\nu}^{\dagger}_{R}$ 
of spin $\frac{1}{2}$, appearing in the $25^{th}$ line of  Table~7 in 
Ref.~\cite{nh2021RPPNP}, or Table~5 in Ref.~\cite{norma2021SQFB} ---
%Table~\ref{Table so13+1.} ---   %(\ref{Table II.})  
are presented in the  left and in the right part of this table, respectively. Table is taken 
from~\cite{normaJMP2015}. 
Families belong to two groups of four families, one ($I$) is a doublet with respect to 
($\vec{\tilde{N}}_{L}$ and  $\vec{\tilde{\tau}}^{1}$) and  a singlet with respect 
to ($\vec{\tilde{N}}_{R}$ and  $\vec{\tilde{\tau}}^{2}$), App.~\ref{generators} 
(Eqs.~(85-88) of Ref.~\cite{nh2021RPPNP}), %\ref{so1+3}, \ref{so42}), 
the other group ($II$) is a singlet with respect to ($\vec{\tilde{N}}_{L}$ and  
$\vec{\tilde{\tau}}^{1}$) and  a doublet with respect to 
($\vec{\tilde{N}}_{R}$ and  $\vec{\tilde{\tau}}^{2}$). 
%, Eqs.~(\ref{so1+3}, \ref{so42}).
All the families follow from the starting one by the application of the operators 
($\tilde{N}^{\pm}_{R,L}$, $\tilde{\tau}^{(2,1)\pm}$). %, Eq.~(\ref{plusminus}).  
The generators ($N^{\pm}_{R,L} $, $\tau^{(2,1)\pm}$) %, Eq.~(\ref{plusminus}),
transform $\hat{u}^{\dagger}_{1R}$ to all the members of one family of the same colour charge. 
The same generators transform equivalently the right handed   neutrino 
$\hat{\nu}^{\dagger}_{1R}$  to all the colourless members of the same family.
\vspace{3mm}}
\label{Table III.}
\end{minipage}
%\end{tiny}
%%\end{small}
 \begin{tabular}{|r|c|c|c|c|c c c c c|}
 \hline
 &&&&&$\tilde{\tau}^{13}$&$\tilde{\tau}^{23}$&$\tilde{N}_{L}^{3}$&$\tilde{N}_{R}^{3}$&
 $\tilde{\tau}^{4}$\\
 \hline
 $I$&$\hat{u}^{c1 \dagger}_{R\,1}$&
   $ \stackrel{03}{(+i)}\,\stackrel{12}{[+]}|\stackrel{56}{[+]}\,\stackrel{78}{(+)} ||
   \stackrel{9 \;10}{(+)}\;\;\stackrel{11\;12}{[-]}\;\;\stackrel{13\;14}{[-]}$ & 
    $\hat{\nu}^{\dagger}_{R\,1}$&
   $ \stackrel{03}{(+i)}\,\stackrel{12}{[+]}|\stackrel{56}{[+]}\,\stackrel{78}{(+)} ||
   \stackrel{9 \;10}{(+)}\;\;\stackrel{11\;12}{(+)}\;\;\stackrel{13\;14}{(+)}$ 
  &$-\frac{1}{2}$&$0$&$-\frac{1}{2}$&$0$&$-\frac{1}{2}$ 
 \\
  $I$&$\hat{u}^{c1 \dagger}_{R\,2}$&
   $ \stackrel{03}{[+i]}\,\stackrel{12}{(+)}|\stackrel{56}{[+]}\,\stackrel{78}{(+)} ||
   \stackrel{9 \;10}{(+)}\;\;\stackrel{11\;12}{[-]}\;\;\stackrel{13\;14}{[-]}$ & 
   $\hat{\nu}^{\dagger}_{R\,2}$&
   $ \stackrel{03}{[+i]}\,\stackrel{12}{(+)}|\stackrel{56}{[+]}\,\stackrel{78}{(+)} ||
   \stackrel{9 \;10}{(+)}\;\;\stackrel{11\;12}{(+)}\;\;\stackrel{13\;14}{(+)}$ 
  &$-\frac{1}{2}$&$0$&$\frac{1}{2}$&$0$&$-\frac{1}{2}$
 \\
  $I$&$\hat{u}^{c1 \dagger}_{R\,3}$&
   $ \stackrel{03}{(+i)}\,\stackrel{12}{[+]}|\stackrel{56}{(+)}\,\stackrel{78}{[+]} ||
   \stackrel{9 \;10}{(+)}\;\;\stackrel{11\;12}{[-]}\;\;\stackrel{13\;14}{[-]}$ & 
    $\hat{\nu}^{\dagger}_{R\,3}$&
   $ \stackrel{03}{(+i)}\,\stackrel{12}{[+]}|\stackrel{56}{(+)}\,\stackrel{78}{[+]} ||
   \stackrel{9 \;10}{(+)}\;\;\stackrel{11\;12}{(+)}\;\;\stackrel{13\;14}{(+)}$ 
  &$\frac{1}{2}$&$0$&$-\frac{1}{2}$&$0$&$-\frac{1}{2}$
 \\
 $I$&$\hat{u}^{c1 \dagger}_{R\,4}$&
  $ \stackrel{03}{[+i]}\,\stackrel{12}{(+)}|\stackrel{56}{(+)}\,\stackrel{78}{[+]} ||
  \stackrel{9 \;10}{(+)}\;\;\stackrel{11\;12}{[-]}\;\;\stackrel{13\;14}{[-]}$ & 
   $\hat{\nu}^{\dagger}_{R\,4}$&
  $ \stackrel{03}{[+i]}\,\stackrel{12}{(+)}|\stackrel{56}{(+)}\,\stackrel{78}{[+]} ||
  \stackrel{9 \;10}{(+)}\;\;\stackrel{11\;12}{(+)}\;\;\stackrel{13\;14}{(+)}$ 
  &$\frac{1}{2}$&$0$&$\frac{1}{2}$&$0$&$-\frac{1}{2}$
  \\
  \hline
  $II$& $\hat{u}^{c1 \dagger}_{R\,5}$&
        $ \stackrel{03}{[+i]}\,\stackrel{12}{[+]}|\stackrel{56}{[+]}\,\stackrel{78}{[+]}||
        \stackrel{9 \;10}{(+)}\;\;\stackrel{11\;12}{[-]}\;\;\stackrel{13\;14}{[-]}$ & 
         $\hat{\nu}^{\dagger}_{R\,5}$&
        $ \stackrel{03}{[+i]}\,\stackrel{12}{[+]}|\stackrel{56}{[+]}\,\stackrel{78}{[+]}|| 
        \stackrel{9 \;10}{(+)}\;\;\stackrel{11\;12}{(+)}\;\;\stackrel{13\;14}{(+)}$ 
        &$0$&$-\frac{1}{2}$&$0$&$-\frac{1}{2}$&$-\frac{1}{2}$
 \\ 
  $II$& $\hat{u}^{c1 \dagger}_{R\,6}$&
      $ \stackrel{03}{(+i)}\,\stackrel{12}{(+)}|\stackrel{56}{[+]}\,\stackrel{78}{[+]}||
      \stackrel{9 \;10}{(+)}\;\;\stackrel{11\;12}{[-]}\;\;\stackrel{13\;14}{[-]}$ & 
    $\hat{\nu}^{\dagger}_{R\,6}$&
      $ \stackrel{03}{(+i)}\,\stackrel{12}{(+)}|\stackrel{56}{[+]}\,\stackrel{78}{[+]}|| 
      \stackrel{9 \;10}{(+)}\;\;\stackrel{11\;12}{(+)}\;\;\stackrel{13\;14}{(+)}$ 
      &$0$&$-\frac{1}{2}$&$0$&$\frac{1}{2}$&$-\frac{1}{2}$
 \\ 
 $II$&$\hat{u}^{c1 \dagger}_{R\,7}$&
 $ \stackrel{03}{[+i]}\,\stackrel{12}{[+]}|\stackrel{56}{(+)}\,\stackrel{78}{(+)}||
 \stackrel{9 \;10}{(+)}\;\;\stackrel{11\;12}{[-]}\;\;\stackrel{13\;14}{[-]}$ & 
      $ \hat{\nu}^{\dagger}_{R\,7}$&
      $ \stackrel{03}{[+i]}\,\stackrel{12}{[+]}|\stackrel{56}{(+)}\,\stackrel{78}{(+)}|| 
      \stackrel{9 \;10}{(+)}\;\;\stackrel{11\;12}{(+)}\;\;\stackrel{13\;14}{(+)}$ 
    &$0$&$\frac{1}{2}$&$0$&$-\frac{1}{2}$&$-\frac{1}{2}$
  \\
   $II$& $\hat{u}^{c1 \dagger}_{R\,8}$&
    $ \stackrel{03}{(+i)}\,\stackrel{12}{(+)}|\stackrel{56}{(+)}\,\stackrel{78}{(+)}||
    \stackrel{9 \;10}{(+)}\;\;\stackrel{11\;12}{[-]}\;\;\stackrel{13\;14}{[-]}$ & 
    $\hat{\nu}^{\dagger}_{R\,8}$&
    $ \stackrel{03}{(+i)}\,\stackrel{12}{(+)}|\stackrel{56}{(+)}\,\stackrel{78}{(+)}|| 
    \stackrel{9 \;10}{(+)}\;\;\stackrel{11\;12}{(+)}\;\;\stackrel{13\;14}{(+)}$ 
    &$0$&$\frac{1}{2}$&$0$&$\frac{1}{2}$&$-\frac{1}{3}$
 \\ 
 \hline 
 \end{tabular}
 \end{tiny}
 \end{center}
 \end{table}

It is the assumption that the eight families from Table~\ref{Table III.} remain massless  
after the break of symmetry from 
$SO(13,1)$ to $SO(7,1) \times SO(6)$, made after   %in the case of $d=(13+1)$.  
we  proved for the toy model~\cite{NHD,ND012} that the break of symmetry 
can leave  some  families of fermions massless, while the rest become massive. 
But we have not yet proven the masslessness  of the $2^{\frac{7+1}{2}-1}$ 
families after the break from $SO(13,1)$ to $SO(7,1) \times SO(6)$. 

The break from the starting symmetry $SO(13,1)$ to 
$SO(7,1) \times SU(3)\times U(1)$ is supposed to  be caused by the appearance 
of the condensate 
of two right handed neutrinos with the family quantum numbers of the upper four 
families, that is of the four  families, which do not contain the three so far observed 
families, at the energy of  $\ge 10^{16}$ GeV.
This condensate is presented in Table~\ref{Table con.}. 
 \begin{table}
 \begin{center}
\begin{minipage}[t]{16.5 cm}
\caption{The condensate of the two right handed neutrinos $\nu_{R}$,  with the quantum numbers
of the $VIII^{th}$ %*****and the $VII^{th}$
family, Table~\ref{Table III.}, coupled to spin zero and belonging to a triplet 
with respect to the generators 
$\tau^{2i}$, is presented, together with its two partners. 
The condensate carries $\vec{\tau}^{1}=0$, $\tau^{23}=1$, 
$\tau^{4}=-1$ and $Q=0=Y$. The triplet carries $\tilde{\tau}^4=-1$, $\tilde{\tau}^{23}=1$
 and $\tilde{N}_{R}^3 = 1$, $\tilde{N}_{L}^3 = 0$,  
$\tilde{Y}=0 $, $\tilde{Q}=0$. 
The family quantum numbers of quarks and leptons are presented in
Table~\ref{Table III.}.  The definition of the operators 
$\vec{\tau}^1, \vec{\tilde{\tau}}^1,  \vec{\tau}^2, \vec{\tilde{\tau}}^2,
\tau^4, \tilde{\tau}^4, N_{R}^3, \tilde{N}_{R}^3,  N_{L}^3, \tilde{N}_{L}^3,
Q, Y, \tilde{Q}, \tilde{Y}$  can be found in App.~\ref{generators} (and in Ref.~\cite{nh2021RPPNP}, Eqs.~(85-88) or in 
Eqs.~(39-41) of Ref.~\cite{norma2021SQFB}).}
\label{Table con.}
\vspace{3mm}
\end{minipage}
%
   % \begin{small}
 \begin{tabular}{|c|c c c c c c c |c c c c c c c|}
 \hline
 state & $S^{03}$& $ S^{12}$ & $\tau^{13}$& $\tau^{23}$ &$\tau^{4}$& $Y$&$Q$&
$\tilde{\tau}^{13}$&
 $\tilde{\tau}^{23}$&$\tilde{\tau}^4$&$\tilde{Y} $& $\tilde{Q}$&$\tilde{N}_{L}^{3}$& 
$\tilde{N}_{R}^{3}$
 \\
 \hline
 ${\bf (|\nu_{1 R}^{VIII}>_{1}\,|\nu_{2 R}^{VIII}>_{2})}$
 & $0$& $0$& $0$& $1$& $-1$ & $0$& $0$& $0$ &$1$& $-1$& $0$& $0$& $0$& $1$\\ 
 \hline
 $ (|\nu_{1 R}^{VIII}>_{1}|e_{2 R}^{VIII}>_{2})$
 & $0$& $0$& $0$& $0$& $-1$ & $-1$& $-1$ & $0$ &$1$& $-1$& $0$& $0$& $0$& $1$\\ 
 $ (|e_{1 R}^{VIII}>_{1}|e_{2 R}^{VIII}>_{2})$
 & $0$& $0$& $0$& $-1$& $-1$ & $-2$& $-2$ & $0$ &$1$& $-1$& $0$& $0$& $0$& $1$\\ 
 \hline 
 \end{tabular}
 %\end{small}
 \end{center}
 \end{table}
%

 % **** 6.1 Reduce text correct it,7.2.2, 7.3.1.\\

% ***  Check Tables 5 and 6\\
% *** 07.11.2021 at 12:15\\
% *** \\

To see how do gravitational fields --- vielbeins and the two spin connection fields, 
the gauge fields of $S^{ab}$ and $\tilde{S}^{ab}$, respectively --- 
contribute to dynamics  of fermion fields and after the electroweak break also to 
the masses of twice four families  and the vector gauge field 
let us rewrite the  fermion part of the action, Eq.~(\ref {wholeaction}),  in the 
way that the fermion action manifests in $d=(3+1)$, that is in the low energy regime
before the electroweak break,
by the {\it standard model} postulated properties, while manifesting the properties 
which make the {\it spin-charge-family} theory a  candidate to go beyond the 
{\it standard model}: \\
$\;\;$ {\bf i.} The spins, handedness, charges and family quantum
numbers of fermions are  % Eqs.~(\ref{so1+3}, \ref{so42}, \ref{so64}), 
determined by the 
Cartan subalgebra of $S^{ab}$ and $\tilde{S}^{ab}$, and the internal space 
of fermions is described by the Clifford "basis vectors" $\hat{b}^{m \dagger}_f$. \\
$\;\;$ {\bf ii.} Couplings
of fermions to the vector gauge fields, which are the superposition of 
gauge fields $\omega^{st}{}_m $, Sect.~\ref{vector3+1}, 
with the space index $m=(0,1,2,3)$ and with 
charges determined by the Cartan subalgebra of $S^{ab}$ and $\tilde{S}^{ab}$ 
($S^{ab} \omega^{cd}{}_{e}= i (\omega^{ad}{}_{e} \eta^{b c} - 
\omega^{bd}{}_{e} \eta^{ac}$) and equivalently for the other two indexes of
$\omega^{cd}{}_{e}$ gauge fields,  manifesting the symmetry of space 
$(d-4)$), and  couplings of fermions to the  scalar gauge fields~\cite{IARD2016,%normaBled2020,
JMP2013,normaJMP2015,%pikan2003,
pikan2006,norma92,norma93,gmdn2008,gn2009,gn2014,IARD2020} with the space 
index $s\ge5$
and the charges determined  by the Cartan subalgebra of $S^{ab}$ 
and $\tilde{S}^{ab}$ (as explained in the case of the vector gauge fields), and 
which are superposition of either $\omega^{st}{}_s $ or 
$\tilde{\omega}^{abt}{}_s $, Sect.~\ref{scalar3+1}
\begin{eqnarray}
\label{faction}
{\mathcal L}_f &=&  \bar{\psi}\gamma^{m} (p_{m}- \sum_{A,i}\; g^{Ai}\tau^{Ai} 
A^{Ai}_{m}) \psi + \nonumber\\
               & &  \{ \sum_{s=7,8}\;  \bar{\psi} \gamma^{s} p_{0s} \; \psi \} +
 \nonumber\\ 
& & \{ \sum_{t=5,6,9,\dots, 14}\;  \bar{\psi} \gamma^{t} p_{0t} \; \psi \}
%+\nonumber\\       & &       {\rm the \;rest}
\,, 
\end{eqnarray}
where $p_{0s} =  p_{s}  - \frac{1}{2}  S^{s' s"} \omega_{s' s" s} - 
                    \frac{1}{2}  \tilde{S}^{ab}   \tilde{\omega}_{ab s}$, 
$p_{0t}   =    p_{t}  - \frac{1}{2}  S^{t' t"} \omega_{t' t" t} - 
                    \frac{1}{2}  \tilde{S}^{ab}   \tilde{\omega}_{ab t}$,                    
with $ m \in (0,1,2,3)$, $s \in (7,8),\, (s',s") \in (5,6,7,8)$, $(a,b)$ (appearing in
 $\tilde{S}^{ab}$) run within  either $ (0,1,2,3)$ or $ (5,6,7,8)$, $t$ runs 
$ \in (5,\dots,14)$, 
$(t',t")$ run either $ \in  (5,6,7,8)$ or $\in (9,10,\dots,14)$. 
The spinor function $\psi$ represents all family members of all the 
$2^{\frac{7+1}{2}-1}=8$ 
families.

$\;\;$ The first line of Eq.~(\ref{faction}) determines in $d=(3+1)$ the kinematics and 
dynamics of fermion fields, coupled to the vector gauge 
fields~\cite{nd2017,normaJMP2015,IARD2016}. 
The vector gauge fields are the superposition of the spin connection fields 
$\omega_{stm}$, $m=(0,1,2,3)$, $(s,t)=(5,6,\cdots,13,14)$, and are the
gauge fields of $S^{st}$, Sect.~\ref{vector3+1}.

% Why gauge vector fields $\tilde{S}^{st}$ are not included?
% See what happens to coupling to the condensate!!

The operators $\tau^{Ai}$ ($\tau^{Ai} = \sum_{a,b} c^{Ai}{ }_{ab}\, 
S^{ab}$, $S^{ab}$ are the generators of the Lorentz transformations in the 
Clifford space of $\gamma^a$'s) are presented in Eqs.~(\ref{so42}, \ref{so64}) 
of App.~\ref{generators}. %Sect.~\ref{examplesClifford}. 
They represent the colour charge, 
$\vec{\tau}^3$, the weak charge, $\vec{\tau}^1$, and the hyper charge,
$Y=\tau^4 + \tau^{23}$, $\tau^4$ is the ''fermion'' charge, originating in 
$SO(6)\subset SO(13,1)$, $\tau^{23}$ belongs together with $\vec{\tau}^1$ of 
$SU(2)_{weak}$ 
to $SO(4)$ ($\subset SO(13+1)$). 

{\it One fermion irreducible representation of the Lorentz group contains}, 
as seen in Table~7 of Ref.~\cite{nh2021RPPNP} or in Table~5 of 
Ref.~\cite{norma2021SQFB},
%\ref{Table so13+1.}, 
{\it quarks and leptons and antiquarks and antileptons}, 
belonging to the first family in Table~\ref{Table III.}. 

Let us repeat again that the 
$SO(7,1)$ subgroup content of the $SO(13,1)$ group is the same for the quarks
and leptons and the same for the antiquarks and antileptons. Quarks distinguish 
from leptons, and antiquarks from antileptons, only in the $SO(6)\subset SO(13,1)$
part, that is in the colour $(\tau^{33},\tau^{38})$ part and in the "fermion" quantum
number $\tau^4$. The quarks distinguish  from antiquarks, and leptons from 
antileptons, in the handedness, in the $SU(2)_{I}\,\rm (weak), \,SU(2)_{II}$, in the colour 
part and in the $\tau^4$ part, explaining 
the relation between handedness and charges of fermions and antifermions, postulated 
in the {\it standard model}%, App.~\ref{appanomalies}%
~\footnote{ 
Ref.~\cite{nh2017} points out that the connection between handedness and charges 
for fermions and antifermions, both appearing in the same irreducible representation, 
explains the triangle anomalies in the {\it standard model} with no need to connect 
''by hand'' the handedness and charges of fermions and antifermions.}.

All the vector gauge fields, which interact with the condensate, presented 
in Table~\ref{Table con.}, become massive, Sect.~\ref{vector3+1}.
The {\it vector gauge fields not interacting  with the condensate --- the weak, 
colour, hyper charge and electromagnetic vector gauge fields --- remain 
massless}, in agreement with by the {\it standard model} assumed gauge fields 
before the electroweak break% of the mass protection
~\footnote{The superposition of the scalar gauge 
fields $\tilde{\omega}^{st}{}_{7}$ and $\tilde{\omega}^{st}{}_{8}$, which at
the electroweak break gain constant values in $d=(3+1)$, bring masses 
to all the vector gauge fields, which couple to these scalar fields.}. 

%
% * Explain why the tilde vector gauge fields do not contribute to the vector 
%   gauge fields!
% 
%  Explanation:
% The superposition of $\tilde{\omega}^{st}{}_{m}$, manifesting as 
% $\tilde{\vec{A}}^{\tilde{1}}_{m}$, $\tilde{A}^{\tilde{Q}}_{m}$,
% $\tilde{A}^{\tilde{Y}}_{m}$, $\tilde{A}^{\tilde{Q}}_{m}$, 
% $\tilde{\vec{A}}^{\tilde{N_{L}}}_{m}$, become massive at the
% electroweak break when all the $\tilde{\omega}^{st}{}_{7,8}$ gain nonzero
% vacuum expectation values.
 
After the electroweak break, caused by the scalar fields, 
%gaining nonzero vacuum expactation values, 
the only conserved charges are the colour and the electromagnetic 
charge $ Q =  \tau^{13} + Y$   ($ Y= \tau^{4} + \tau^{23}$).
All the rest interact with the scalar fields of the constant value.

\vspace{2mm}

 $\;\;$ The second line of Eq.~(\ref{faction}) is the mass term, responsible 
in $d=(3+1)$  for the masses of fermions and of the weak gauge field 
(originating in spin connection fields $\omega^{s t}{}_{m}$). 
The interaction of fermions with the scalar fields with the space index 
$s=(7,8)$ (to these scalar fields particular superposition of the spin connection 
fields $\omega^{a b}{}_{s}$ and all the superposition of
$\tilde{\omega}^{a b}{}_{s}$  with the space index $s=(7,8)$ and
$(a,b)=(0,1,2,3)$ or $(a,b)=(5,6,7,8)$ contribute),  which gain the constant  
values in $d=(3+1)$, makes fermions and antifermions massive. 
% These scalar fields make massive also the weak vector gauge fields. 

{\it The scalar fields, presented in the second line of Eq.~(\ref{faction}), are
in the {\t standard model} interpreted as the higgs and  the Yukawa couplings}, 
Sect.~\ref{scalar3+1}, predicting in the {\it spin-charge-family} theory
that there must exist several scalar fields~\footnote{The requirement of the 
{\it standard model} that there exist the Yukawa couplings, speaks by itself
that there must exist several scalar fields explaining the Yukawa couplings.}. 

%**24.02.2021 at 12:00\\

% **SEE if you would put this in Subsubsections!! shorten for here

%****************************

These  scalar gauge fields split into two groups of scalar fields. 
One group of two triplets and three singlets manifests the symmetry 
$\widetilde{SU}(2)_{(\widetilde{SO}(3,1), L)}$ 
$\times \widetilde{SU}(2)_{(\widetilde{SO}(4), L)}$ $\times U(1)$.  The 
other group of another two triplets and the same three singlets manifests the 
symmetry  $\widetilde{SU}(2)_{(\widetilde{SO}(3,1), R)}$ 
$\times \widetilde{SU}(2)_{(\widetilde{SO}(4), R)}$ $\times U(1)$. 

The three  $U(1)$ singlet scalar gauge fields are superposition of  
$\omega_{s' t' s}$, $s=(7,8)$, $(s',t')=(5,6,\cdots,14)$, with the  sums of 
$S^{s' t'}$ arranged into 
superposition of $\tau^{13}$, $\tau^{23}$ and $\tau^4$. The three triplets
interact with both groups of quarks and leptons and antiquarks and 
antileptons~\cite{mdn2006,gmdn2007,gmdn2008,gn2009,gn2014,NA2018,%
NH2017newdata}.

Families of fermions from Table~\ref{Table III.}, interacting with these scalar 
fields, split  as well into two groups of four families,  each of these two groups are 
coupled to one of the two groups of scalar triplets while all eight families 
couple to the same three singlets. The scalar gauge fields, manifesting 
$\widetilde{SU}(2)_{L,R} \times \widetilde{SU}(2)_{L,R}$, are the superposition 
of the gauge fields 
$\tilde{\omega}_{ab s}$, $s=(7,8), (a,b) = $ either $(0,1,2,3)$ or $(5,6,7,8)$, 
manifesting as twice two triplets. 
%the symmetry  of  $SO(3,1) \times SO(4)$ included in $SO(7,1)$.

%**Below should go to subsubsection

%
\subsection{Vector and scalar gauge fields before electroweak break}
\label{vectorscalar3+1}
%Short report, refer to nh2021RPPNP Include even Clifford representations
The second line of Eq.~(\ref{wholeaction}) represents the action for the gauge fields
$A_{gf}$
\begin{eqnarray}
A_{gf}&=& \int \; d^dx \; E\; (\alpha \,R + \tilde{\alpha} \, \tilde{R})\,,
\nonumber\\                
R &=&  \frac{1}{2} \, \{ f^{\alpha [ a} f^{\beta b ]} \;(\omega_{a b \alpha, \beta} 
- \omega_{c a \alpha}\,\omega^{c}{}_{b \beta}) \} + h.c. \,, \nonumber \\
\tilde{R}  &=&  \frac{1}{2} \, \{ f^{\alpha [ a} f^{\beta b ]} 
\;(\tilde{\omega}_{a b \alpha,\beta} - \tilde{\omega}_{c a \alpha} \,
\tilde{\omega}^{c}{}_{b \beta})\} + h.c.\,.               
\label{gfaction}
\end{eqnarray}
It is proven in Ref.~\cite{nd2017} that the vector and the scalar 
gauge fields manifest in $d=(3+1)$, after the break of the starting symmetry, as 
the superposition of spin connection fields, when the space $(d-4)$ manifest the
assumed symmetry.
$f^{\beta}{}_{a}$  and $e^{a}{}_{\alpha}$ are vielbeins and inverted  vielbeins 
respectively, 
$e^{a}{}_{\alpha}f^{\beta}{}_{a} =\delta^{\beta}_{\alpha}$, 
$e^{a}{}_{\alpha}f^{\alpha}{}_{b}= \delta^{a}_{b}$,
%\end{eqnarray}
%
$E =det(e^{a}{}_{\alpha})$.

% *** 14.11.2021 at 14:30

Varying the action of Eq.~(\ref{gfaction})  with respect to the spin 
connection fields the expression for the spin connection fields $\omega_{ab}{}^e$ 
follows
\begin{eqnarray}
\label{omegaabe}
\omega_{ab}{}^{e} &=& 
 \frac{1}{2E} \{   e^{e}{}_{\alpha}\,\partial_\beta(Ef^{\alpha}{}_{[a} f^\beta{}_{b]} )
      - e_{a\alpha}\,\partial_\beta(Ef^{\alpha}{}_{[b}f^{\beta e]})
 % \nonumber\\                  & &  \qquad\qquad  
{} - e_{b\alpha} \partial_\beta (Ef^{\alpha [e} f^\beta{}_{a]})\}
                     \nonumber\\
                  &+& \frac{1}{4}   \{\bar{\Psi} (\gamma^e \,S_{ab} - 
 \gamma_{[a}  S_{b]}{}^{e} )\Psi \}  \nonumber\\
                  &-& \frac{1}{d-2}  
   \{ \delta^e_{a} [
\frac{1}{E}\,e^d{}_{\alpha} \partial_{\beta}
             (Ef^{\alpha}{}_{[d}f^{\beta}{}_{b]})
                        + \bar{\Psi} \gamma_d  S^{d}{}_{b} \,\Psi ] 
\nonumber\\  % & & \qquad 
{}  &-& \delta^{e}_{b} [\frac{1}{E} e^{d}{}_{\alpha} \partial_{\beta}
             (Ef^{\alpha}{}_{[d}f^{\beta}{}_{a]} )
            + \bar{\Psi} \gamma_{d}  S^{d}{}_{a}\, \Psi ]\}\,. %\label{omegas} 
                        \end{eqnarray}
Replacing $S^{ab}$ in Eq.~(\ref{omegaabe}) with $\tilde{S}^{ab}$, the  expression
for the spin connection fields  $\tilde{\omega}_{ab}{}^{e}$ follows.

If there are no spinors (fermions) present, $\Psi =0$, then either 
$\omega_{ab}{}^{e} $ or $\tilde{\omega}_{ab}{}^{e} $ are uniquely expressed 
with the vielbeins.

Spin connection fields $\omega^{ab}{}_{e} $ represent  vector gauge fields 
to the corresponding  fermion fields if index $e$ is $m=(0,1,2,3)$. If $e\ge 5$
the spin connection fields manifest in $d=(3+1)$ as scalar gauge fields.

 It is proven in Ref.~\cite{nd2017}~%
\footnote{We presented in 
Ref.~\cite{nd2017} the proof, that the vielbeins $f^\sigma{}_m$ (Einstein index 
$\sigma \ge 5$, $m=0,1,2,3$)  lead in $d=(3+1)$ to the vector gauge fields, 
which are the superposition of the spin connection fields $\omega_{st m}$: 
 $f^\sigma{}_m=  \sum_{A}\,  \vec{A}^{A}_m$ 
$\vec{\tau}^{A \sigma}{}_{\tau}\, x^{\tau}$, 
with $A^{Ai}_{m}=\sum_{s,t} c^{Ai}{}_{st}\, \omega^{st}{}_{m}$, when 
the metric in $(d-4)$, $g_{\sigma \tau}$, is invariant under the coordinate 
transformations $x^{\sigma'} = x^{\sigma} + \sum_{A,i,s,t} \varepsilon^{A i}\,
(x^{m})\,c^{A i}{}_{st}$  $E^{\sigma s t} (x^{\tau})$ and 
$\sum_{s,t} c^{A i}{}_{st} \, E^{\sigma s t} = \tau^{A i \sigma}$, while  
$\tau^{A i \sigma}$  solves the Killing equation: 
$D_{\sigma}\, \tau^{A i}_{\tau} + D_{\tau} \tau^{A i}_{\sigma} =0\,$
($D_{\sigma}\,  \tau^{A i}_{\tau} = \partial_{\sigma}\, \tau^{A i}_{\tau} - 
\Gamma^{\tau'}_{\tau \sigma} \tau^{Ai}_{\tau'})$. And similarly also for 
the scalar gauge fields.} that in spaces with the desired symmetry 
%allowing the below expression 
the vielbein can be expressed with the gauge fields,
% if only one of the two spin connection fields are present
%
\begin{eqnarray}
\label{fmagen} 
f^{\sigma}{}_{m}&=& \sum_{A}\,\vec{\tau}^{A\sigma}\, \vec{A}^{A}_{m}\,, \nonumber\\
\tau^{Ai \sigma} &=&  \sum_{st}\, c^{Ai}{}_{st}\,  (e_{s  \tau}\,
 f^{\sigma}{}_{t} - e_{t  \tau}\,
f^{\sigma}{}_{s}) x^{\tau}\, , \nonumber\\
A^{Ai}_{m}&=& \sum_{st} \,c^{Ai}{}_{st} \, \omega^{st}{}_{m}\,, 
\nonumber\\
\tau^{Ai} &=& \sum_{st}\, c^{Ai}{}_{st}\,S^{st} \,
\,,\nonumber\\
\{\tau^{Ai}, \tau^{Bj}\}_- &=& i \delta^{AB} f^{Aijk} \tau^{Ak}\,. 
\end{eqnarray}
The vector gauge fields $ A^{Ai}_{m}$ of  $\tau^{Ai}$  represent in the 
{\it spin-charge-family} theory all the observed gauge fields, as well as the 
additional non observed vector gauge fields, which interacting with the 
condensate gain heavy masses.
%*26.02.2021. at 20:45\\

The scalar (gauge) fields, carrying the space index $s=(5,6,\dots,d)$,  offer 
in the {\it spin-charge-family} for $s=(7,8)$ the explanation for the origin of 
the Higgs's scalar and the Yukawa couplings of the {\it standard model}, 
while scalars with the space index $s=(9,10,\dots, 14)$ offer  the 
explanation for the proton decay, as well as for the matter/antimatter 
asymmetry in the universe. 

In the scalar gauge fields besides $\omega^{st}{}_{s '}$ also 
$\tilde{\omega}^{ab}{}_{s}$ contribute.

The explicit expressions for $c^{Ai}{ }_{ab}$, and correspondingly for $\tau^{Ai}$, 
and $ A^{Ai}_{a}$, are written in Sects.~%\ref{vector3+1} and~\ref{scalar3+1}. 
4.2.1. and 4.2.2 of Ref.~\cite{nh2021RPPNP}.

% and in App.~\ref{}.

\subsubsection{Vector gauge fields}
\label{vector3+1}
%
%Short report, refer to nh2021RPPNP    Include even Clifford representations

All the vector gauge fields are in the {\it spin-charge-family }theory expressible with 
the spin connection fields  $\omega_{stm}$ 
% before the break of symmetries in a simple way
as
\begin{eqnarray}
 A^{Ai}_{m} &=& \sum_{s,t} \;c^{Ai}{ }_{st} \; \omega^{st}{}_{m}\,, 
 \label{vector}
 \end{eqnarray}
with $\sum_{A,i} \tau^{Ai} \,A^{Ai}_{m} = \sum^*_{a,b} S^{ab} 
\omega^{ab}{}_{m}$,${}^*$ means that summation runs over $(a,b)$  
respecting the symmetry $SO(7,1)\times SU(3)\times U(1)$,  with $SO(7,1)$
breaking further to $SO(3,1)\times SU(2)_I \times  SU(2)_{II} $.

The vector gauge fields are namely analysed from the
point of view of the possibly observed fields in $d=(3+1)$ space: besides
gravity, the colour $SU(3)$, the weak $SU(2)_{I}$, the second $SU(2)_{II}$ and
the  $U(1)_{\tau^4}$ - the vector gauge field of the  "fermion" quantum number 
$\tau^4$.

Due to the interaction with the condensate the second $SU(2)_{II}$ (one superposition 
of the third component of $SU(2)_{II}$ and of the  $U(1)_{\tau^4}$ vector gauge fields
and the rest two components of the $SU(2)_{II}$ vector gauge field) become
massive, while the colour $SU(3)$, the weak $SU(2)_{I}$,  the second superposition of 
the third component of $SU(2)_{II}$ and the  $U(1)_{\tau^4}$, forming the hyper charge
vector gauge field, remain massless.
That is:
All the vector gauge fields, as well as the scalar gauge fields of  $S^{ab}$ and 
of $\tilde{S}^{ab}$, 
% $\vec{A}^{2}_{m}$ and $A^{4}_{m}$ gauge fields and 
% scalar gauge fields of $S^{ab}$, 
which interact with the condensate,  
% the gauge fields $\vec{\tilde{A}}^{Ai}_{m}$ and scalar gauge fields 
% $\vec{\tilde{A}}^{Ai}_{s}, s\ge 5$, which couple to the condensate, 
become massive.

The effective action for all the massless vector gauge fields, the gauge fields 
which  do not interact with the condensate and remain therefore massless,
before the electroweak break,  
equal to  $\int \,d^{4} x \, \{ - \frac{1}{4} \,
 F^{A i}{}_{m n} $ $F^{A i m n}\,\} $, with  the structure constants  $f^{Aijk}$
concerning the colour $SU(3)$, weak SU(2) and hyper charge $U(1)$ 
groups~\cite{nd2017}.

All these relations are valid as long as spinors and vector gauge fields  are weak 
fields in comparison with the fields which force $(d-4)$ space to be (almost) curled, 
Ref.~\cite{TDN}. When 
all these fields, with the scalar gauge fields included,  start to be comparable with 
the fields (spinors or scalars), which determine the symmetry of $(d-4)$ space, 
the symmetry of the whole space changes.

The electroweak break, caused by the constant (non zero vacuum expectation) 
values of the scalar gauge fields, carrying the space index $s=(7,8)$, makes 
the weak and the hyper charge gauge fields massive. The only vector gauge 
fields which remain massless are, besides the  gravity, the 
electromagnetic and the colour vector gauge fields --- the observed three massless
gauge fields.

% *Check what breaks $SO(6)$! Mention App.\\ 
% *Can this do the condensate through $\tau^4$?

 %
\subsubsection{Scalar gauge fields in $d=(3+1)$}
\label{scalar3+1}
%
%Short report, refer to nh2021RPPNP

The starting action of the {\it spin-charge-family} theory offers scalar fields of two
kinds:\\ 
{\bf a.} Scalar fields, taking care of the masses of quarks and leptons have  the space 
index $s=(7,8)$ and carry with respect to this space index the weak charge $\tau^{13} =\pm \frac{1}{2}$ and the hyper charge $Y=\mp \frac{1}{2}$, Table~\ref{Table doublets.}, Eq.~(\ref{checktau13Y}). 
With respect to the index  $Ai$, determined by the relation $\tau^{Ai}= 
\sum_{ab}  c^{Ai}{}_{ab} S^{ab}$ 
and $\tilde{\tau}^{Ai}= \sum_{ab}  c^{Ai}{}_{ab} \tilde{S}^{ab}$, that is with
respect to $S^{ab}$ and $\tilde{S}^{ab}$, they carry charges and family 
charges in adjoint representations. \\
%
% (${\cal S}^{ab} \, A^{d\dots e \dots g} = i \,(\eta^{ae} \,A^{d\dots b \dots g} - 
%\eta^{be}\,A^{d\dots a \dots g} )$. \\
{\bf b.}
There are in the starting action of the {\it spin-charge-family} theory,
 Eq.~(\ref{wholeaction}), scalar fields, which transform antileptons and 
antiquarks into quarks and leptons and back. They carry space index 
$s=(9,10,\dots,14)$, They are with respect to the space index colour 
triplets and antitriplets, while they carry charges  $\tau^{Ai}$ and
$\tilde{\tau}^{Ai}$ in adjoint representations. 

Following Refs.~\cite{normaJMP2015,IARD2020,nh2021RPPNP} I shall 
review both kinds of scalar fields.%
~\footnote{
Let us demonstrate how do the  infinitesimal generators ${\cal S}^{ab}$ 
apply on the spin connections fields $\omega_{bd e}$ ($= f^{\alpha}{}_{e}\, $ 
$\omega_{bd \alpha}$) and $\tilde{\omega}_{\tilde{b} \tilde{d} e}$ 
($= f^{\alpha}{}_{e}\,$ $\tilde{\omega}_{\tilde{b} \tilde{d} \alpha}$), on 
either the space index $e$ or any of the
indices $(b,d,\tilde{b},\tilde{d})$ 
${\cal S}^{ab} \, A^{d\dots e \dots g} = i \,(\eta^{ae} \,A^{d\dots b \dots g} - 
\eta^{be}\,A^{d\dots a \dots g} )$ 
(Section~IV. and Appendix~B in Ref.~\cite{normaJMP2015}).}

\vspace{4mm}

{\it {\bf a. Scalar gauge fields determining scalar higgs and Yukawa couplings}}

\vspace{4mm}

Making a choice of the scalar index equal to $s=(7,8)$ (the choice of $(s=5,6)$ 
would also work) and allowing all superposition of $\tilde{\omega}_{\tilde{a}\tilde{b}s}$, 
while with respect to $\omega_{abs}$ only the  superposition representing 
the scalar gauge fields $A^{Q}_{s}$,  $A^{Y}_{s}$ and $A^{4}_s$, 
$s=(7,8)$ (or any three superposition of these three scalar fields) may 
contribute. 

Let us use the common notation  $ A^{Ai}_{s}$ for all the scalar 
gauge fields with $s=(7,8)$, independently of whether  they originate in 
$\omega_{abs}$ --- in this case $ Ai$ $=(Q$,$Y,\tau^4$) --- or in 
$\tilde{\omega}_{\tilde{a}\tilde{b}s}$. All these gauge fields  contribute to 
the masses of quarks and leptons and antiquarks and antileptons after 
gaining constant values (nonzero vacuum expectation values).
\begin{eqnarray}
\label{commonAi}
 A^{Ai}_{s} &{\rm represents}& (\,A^{Q}_{s}\,,A^{Y}_{s}\,, A^{4}_{s}\,, 
 % \tilde{\omega}_{\tilde{a} \tilde{b} s }\,,
 \vec{\tilde{A}}^{\tilde{1}}_{s}\,, 
 \vec{\tilde{A}}^{\tilde{N}_{\tilde{L}}}_{s}\,, \vec{\tilde{A}}^{\tilde{2}}_{s}\,, 
 \vec{\tilde{A}}^{\tilde{N}_{\tilde{R}}}_{s}\,)\,,\nonumber\\
\tau^{Ai} &{\rm represents}& (Q,\,Y,\,\tau^4, \,\vec{\tilde{\tau}}^{1},\,
 \vec{\tilde{N}}_{L},\,\vec{\tilde{\tau}}^{2},\,\vec{\tilde{N}}_{R})\,.
\end{eqnarray}
Here $\tau^{Ai}$ represent all the operators which apply on fermions.
These scalars with the space index $s=(7,8)$, they are  scalar gauge fields 
of the generators $\tau^{Ai}$ and 
$\tilde{\tau}^{Ai}$, %(Eqs.~(\ref{so64} - \ref{so42tilde})), 
are expressible in terms of the spin connection fields, App.~\ref{generators} (Ref.~\cite{normaJMP2015}, Eqs.~(10, 22, A8, A9)).

% ***09.11.2021 at 19:00 7606 \cite{nh2021RPPNP} Eq.(108)

All the scalar fields with the 
space index $(7,8)$ carry with respect to this space index the weak and the 
hyper charge ($\mp \frac{1}{2}$, $\pm \frac{1}{2}$), respectively, all  
having therefore properties as required for the higgs in the {\it standard model}. 

To make the scalar fields the eigenstates of $\tau^{13}= 
\frac{1}{2}({\cal S}^{56} - {\cal S}^{78})$ and to check their properties with
respect to $Y$ $(= \tau^{4} + \tau^{23} =(\frac{1}{2} 
({\cal S}^{56} +{\cal S}^{78})  - \frac{1}{3} ({\cal S}^{9\,10}
 +  {\cal S}^{11\,12} +{\cal S}^{13\,14}))$ and $Q$ $( =  \tau^{13} + Y)$
 we need to apply the operators $\tau^{13}$, $Y$  and $Q$  on the scalar fields 
with the space index $s=(7,8)$, taking into account the relation
${\cal S}^{ab} \, A^{d\dots e \dots g} = i \,(\eta^{ae} \,A^{d\dots b \dots g} - 
\eta^{be}\,A^{d\dots a \dots g} )$.  
%(Section~IV. and Appendix~B in Ref.~\cite{normaJMP2015}).

Let us  rewrite the second line of Eq.~(\ref{faction}), paying no attention to the 
momentum $p_{s}\,, s\in(5,\dots,8)$, when having in mind the lowest energy solutions manifesting at low energies.
\begin{eqnarray}
\label{eigentau1tau2}
 & &\sum_{s=(7,8), A,i}\, \bar{\psi} \,\gamma^s\, ( - \tau^{Ai} \,A^{Ai}_{s}\,)
\,\psi = \nonumber\\
 & &-\sum_{A,i} \bar{\psi}\,\{\,\stackrel{78}{(+)}\, \tau^{Ai} \,(A^{Ai}_{7} - i   
 \,A^{Ai}_{8})\, + \stackrel{78}{(-)}(\tau^{Ai} \,(A^{Ai}_{7} + i \,A^{Ai}_{8})\,\}
\,\psi\,,  \nonumber\\
 & &\stackrel{78}{(\pm)} = \frac{1}{2}\, (\gamma^{7} \pm \,i \, \gamma^{8}\,)
\,,\quad  A^{Ai}_{\scriptscriptstyle{\stackrel{78}{(\pm)}}}: = (A^{Ai}_7 \,\mp i\, 
A^{Ai}_8)\,,
\end{eqnarray}
with the summation over $A$ and $i$ performed, with $A^{Ai}_s$ representing the 
scalar fields ($A^{Q}_{s}$, $A^{Y}_{s}$, $A^{4}_{s}$) determined by 
$\omega_{s',s'',s}\;$, as well as  ($\tilde{A}^{\tilde{4}}_{s}$, 
$\vec{\tilde{A}}^{\tilde{1}}_{s}$, $\vec{\tilde{A}}^{\tilde{2}}_{s}$, 
$\vec{\tilde{A}}^{\tilde{N}_{R}}_{s}$ and $\vec{\tilde{A}}^{\tilde{N}_{L}}_{s}$),
determined by $\tilde{\omega}_{a,b,s}\,, s=(7,8)$.

The application of the operators $ \tau^{13}$, $Y$ and $Q$
on the scalar fields ($A^{Ai}_{7}\mp i\,A^{Ai}_{8})$ with respect to the space 
index $s=(7,8)$,  gives  
\begin{eqnarray}
\label{checktau13Y}
\tau^{13}\,(A^{Ai}_7 \,\mp i\, A^{Ai}_8)&=& \pm \,\frac{1}{2}\,(A^{Ai}_7 \,
\mp i\, A^{Ai}_8)\,,\nonumber\\
Y\,(A^{Ai}_7 \,\mp i\, A^{Ai}_8)&=& \mp \,\frac{1}{2}\,(A^{Ai}_7 \,
\mp i\, A^{Ai}_8)\,,\nonumber\\
Q\,(A^{Ai}_7 \,\mp i\, A^{Ai}_8)&=& 0\,.
\end{eqnarray}
Since  $ \tau^{4}$, $Y$, $\tau^{13}$ and $\tau^{1 +},  \tau^{1 -}$ give zero if 
applied on  ($A^{Q}_{s}$, $A^{Y}_{s}$ and $A^{4}_{s}$) (with  respect to the 
quantum numbers ($Q, Y, \tau^4$)),  and since $Y, Q, \tau^4$ and $\tau^{13}$ 
commute with the family quantum numbers, one sees that the scalar fields $A^{Ai}_{s}$ 
( =($A^{Q}_{s}$, $A^{Y}_{s}$, $A^{Y'}_{s}$, 
$\tilde{A}^{\tilde{4}}_{s}$, $\tilde{A}^{\tilde{Q}}_{s}$, 
$\vec{\tilde{A}}^{\tilde{1}}_{s}$, 
$\vec{\tilde{A}}^{\tilde{2}}_{s}$, $\vec{\tilde{A}}^{\tilde{N}_{R}}_{s}$,  
$\vec{\tilde{A}}^{\tilde{N}_{L}}_{s}$)), $s=(7,8)$, rewritten as  
$A^{Ai}_{\scriptscriptstyle{\stackrel{78}{(\pm)}}} $ 
$= (A^{Ai}_7 \,\mp i\, A^{Ai}_8)\,$,
are eigenstates of $\tau^{13}$ and $Y$, having the quantum numbers of the 
{\it standard model} Higgs's  scalar.

These superposition of $A^{Ai}_{\scriptscriptstyle{\stackrel{78}{(\pm)}}}$ are 
presented in 
Table~\ref{Table doublets.} as two doublets with respect to the weak charge 
${\cal \tau}^{13}$,  with the eigenvalue of ${\cal \tau}^{23}$  (the second 
$SU(2)_{II}$ charge) 
equal to either $-\frac{1}{2}$ or $+\frac{1}{2}$, respectively. 
\begin{table}
\begin{center}
\begin{minipage}[t]{16.5 cm}
\caption{The two scalar weak doublets, one with $ {\cal \tau}^{23}=- \frac{1}{2}$  
and the other with $ {\cal \tau}^{23}=+ \frac{1}{2}$, both with the "fermion"
quantum number ${\cal \tau}^{4}$ $=0$, are presented. 
In this table all the scalar fields carry besides the quantum numbers determined by 
the space index also  the quantum numbers $A$ and $i$ from 
Eq.~(\ref{commonAi}). The table is taken from Ref.~\cite{normaJMP2015}.}
\label{Table doublets.}
\vspace{3mm}
\end{minipage}  
%% \begin{small}
{ \begin{tabular}{|c|c| c c c c r|}
 \hline
 name & superposition & ${\cal \tau}^{13}$& $ {\cal \tau}^{23}$ & spin& ${\cal \tau}^{4}$& $ Q$\\
 \hline
 $A^{Ai}_{\scriptscriptstyle{\stackrel{78}{(-)}}}$ & $A^{Ai}_{7}+iA^{Ai}_{8}$& $+
\frac{1}{2}$& 
 $-\frac{1}{2}$& 0&0& 0\\
 $A^{Ai}_{\scriptscriptstyle{\stackrel{56}{(-)}}}$ & $A^{Ai}_{5}+iA^{Ai}_{6}$& 
$-\frac{1}{2}$& 
 $-\frac{1}{2}$& 0&0& -1\\
 \hline 
$A^{Ai}_{\scriptscriptstyle{\stackrel{78}{(+)}}}$ & $A^{Ai}_{7}-iA^{Ai}_{8}$& 
$-\frac{1}{2}$& 
$+\frac{1}{2}$& 0&0& 0\\
$A^{Ai}_{\scriptscriptstyle{\stackrel{56}{(+)}}}$ & $A^{Ai}_{5}-iA^{Ai}_{6}$& $+
\frac{1}{2}$& 
$+\frac{1}{2}$& 0& 0&+1\\ 
\hline
\end{tabular}
}
 %%\end{small}
 \end{center}
%\label{Table doublets.}
 \end{table}

It is not difficult to show that the scalar fields 
$A^{Ai}_{\scriptscriptstyle{\stackrel{78}{(\pm)}}}$  
are {\it triplets} as the gauge fields of the  family quantum numbers 
($\vec{\tilde{N}}_{R}, \,$ $\vec{\tilde{N}}_{L},\,$ $ \vec{\tilde{\tau}}^{2},\,$ 
$\vec{\tilde{\tau}}^{1}$
 %Eqs.~(\ref{so1+3}, \ref{so42}, \ref{bosonspin0})) 
 or singlets as the gauge fields of 
$Q=\tau^{13}+Y, \,Q'= -\tan^{2}\vartheta_{1} Y$ $ + \tau^{13} $ and
 $Y' = -\tan^2 \vartheta_{2} \tau^{4} + \tau^{23}$. %~%

Table~\ref{Table III.} represents two groups of four families. It is not difficult 
to see that $ \tilde{N}^{\pm}_{L}$ and $ \tilde{\tau}^{1 \pm}$ transform the 
first four families among themselves, leaving the second group of four families 
untouched, while $ \tilde{N}^{\pm}_{R}$ and $ \tilde{\tau}^{2 \pm}$  do not 
influence  the first four families and transform the second four families among 
themselves. All the scalar fields with $s=(7,8)$ "dress" the right handed quarks 
and leptons with the hyper charge and the weak charge so that they manifest  
charges of the left handed partners. 

The  mass matrices $4 \times 4$, representing the application of the scalar gauge 
fields on fermions of each of the two groups,
have the symmetry $SU(2)\times SU(2)\times U(1)$ of the form as presented in Eq.~(\ref{M0})~\footnote{%
The symmetry $SU(2)\times SU(2)\times U(1)$  of the mass matrices, 
Eq.~(\ref{M0}), is expected to remain in all loop corrections~\cite{NA2018}.}.
The  influence of scalar fields on the masses of quarks and leptons 
depends on the coupling constants  and the masses of the scalar fields, determining 
parameters of the mass matrix
\begin{small}
 \begin{equation}
 \label{M0}
 {\cal M}^{\alpha} = \begin{pmatrix} 
 - a_1 - a & e     &   d & b\\ 
 e*     & - a_2 - a &   b & d\\ 
 d*     & b*     & a_2 - a & e\\
 b*     &  d*    & e*   & a_1 - a
 \end{pmatrix}^{\alpha}\,,
 \end{equation}
 \end{small}
%  l 2100
with $\alpha$ representing family members --- quarks and leptons~%
\cite{mdn2006,gmdn2007,gmdn2008,gn2014,NH2017newdata}. 
In Subsect.~\ref{predictions} the predictions of the {\it spin-charge-family}
theory following from the symmetry of mass matrices of Eq.~(\ref{M0})
are discussed. 

The {\it spin-charge-family} theory treats quarks and leptons in equivalent 
way. The differences among family members occur due to the scalar fields
($Q \cdot A^{Q}_{\scriptscriptstyle{\stackrel{78}{(\pm)}}},
Y \cdot A^{Q}_{\scriptscriptstyle{\stackrel{78}{(\pm)}}},
\tau^4 \cdot A^{4}_{\scriptscriptstyle{\stackrel{78}{(\pm)}}}$)~%.
\cite{gn2014,NH2017newdata}.

Twice four families of Table~\ref{Table III.}, with the two groups of two 
triplets applying each on one of the two groups of four families and one group 
of three singlets applying on all eight families, {\bf i.} offer the explanation for  the 
appearance of the Higgs's scalar and Yukawa couplings of the observed three
families, predicting the fourth family to the observed three families and 
several scalar fields, {\bf ii.}  predict that the stable of the additional four 
families with much higher masses that the lower four families contributes to the 
{\it dark matter}.

\vspace{4mm}

{\bf b.} {\it {\bf Scalar gauge fields causing transitions from antileptons and antiquarks 
 into quarks and leptons}}~\cite{n2014matterantimatter}

\vspace{4mm}

Besides the scalar fields with the space index $s=(7,8)$ which manifest
in $d=(3+1)$ as scalar gauge fields with the weak and hyper charge 
$\pm \frac{1}{2}$ and $\mp \frac{1}{2}$, respectively, and which gaining at low 
energies constant values cause masses of families of
quarks and leptons and of the weak gauge field, there are in the starting
action, Eqs.~(\ref{wholeaction}, \ref{faction}), additional scalar gauge
fields with the space index $t=(9,10,11,12,13,14)$. They are with respect
to the space index $t$ either triplets or antitriplets causing transitions from 
antileptons into quarks and from antiquarks into quarks and back. 
These scalar fields are in Eq.~(\ref{faction}) presented in the third line. 

These scalar fields are offering the explanation for the matter/antimater 
asymmetry in the universe, and might be responsible for proton decay and 
lepton number nonconservation. The reader is kindly ask to read the article~\cite{n2014matterantimatter}, for a short review one can see the
Refs.~\cite{IARD2016,nh2021RPPNP}.

\subsection{Predictions of {\it spin-charge-family} theory}
\label{predictions}

Let me say that the fact that the simple starting action, 
Eq.~(\ref{wholeaction}) --- in which fermions interact with gravity only (the 
vielbeins and the two kinds of the spin connection fields), while the internal 
spaces  of fermions and bosons are describable by the ''basis vectors'' which
are superposition of odd or even products of Clifford algebra operators 
$\gamma^a$'s, respectively --- offers the explanation for all the assumptions 
of the {\it standard model}  and for the second quantized postulates for 
fermions and bosons,  while unifying all the so far known forces, with 
gravity included, predicting new vector gauge fields, new scalar gauge fields 
and new families of fermions, gives a hope that the {\it spin-charge-family} 
theory is offering the right next step beyond the {\it standard model}.

\vspace{2mm}

{\bf i.} The existence of the lower group of four families predicts the 
fourth family to the observed three, which should be seen in next 
experiments. The masses of quarks of these four families are 
determined by several scalar fields, all with the properties of the scalar 
higgs, some of them of which might also be observed.

The symmetry~\cite{NA2018,gn2014}, Eq.~(\ref{M0}), and the values 
of mass matrices 
of the lower four families are determined with two triplet scalar fields,  
$ \vec{\tilde{A}}^{\tilde{1}}_{\stackrel{78}{(\pm)}}$
and $\vec{\tilde{A}}^{\tilde{N}_{\tilde{L}}}_{\stackrel{78}{(\pm)}}$, 
and  three singlet scalar fields, $A^{Q}_{\stackrel{78}{(\pm)}}$, 
$A^{Y}_{\stackrel{78}{(\pm)}}$, $ A^{4}_{\stackrel{78}{(\pm)}}$,
Eq.~(\ref{commonAi}), explaining  the Higgs's scalar and  Yukawa 
couplings of the {\it standard model}, 
Refs.~\cite{gn2014,normaJMP2015,IARD2016,NH2017newdata,normaBled2020} 
and references therein.

Any accurate $3\times 3$ submatrix of the $4 \times 4$ unitary matrix 
determines the $4 \times 4$ matrix uniquely. Since neither the quark and 
(in particular) nor the lepton $3\times 3$ mixing matrix are measured 
accurately enough to be able to determine three complex phases of the 
$4 \times 4$ mixing matrix, we assume (what also simplifies the 
numerical procedure)~\cite{mdn2006,gmdn2007,gmdn2008,%
gn2009,gn2014} that the mass matrices are symmetric and real 
and correspondingly the mixing matrices are orthogonal. We fitted the 
$6$ free parameters of each family member mass matrix, Eq.~(\ref{M0}),  
to twice three measured masses ($6$) of each pair of either quarks or 
leptons and to the $6$ (from the experimental data extracted) parameters 
of the corresponding $4 \times 4$ mixing matrix.

I present here the old results for quarks only, taken from  Refs.~\cite{gn2014}.  
The accuracy of the experimental data for leptons are not yet large 
enough that would allow any meaningful prediction~%
\footnote{The numerical procedure, explained in the paper~\cite{gn2014}, 
to fit free parameters of the mass matrices to the experimental data within 
the experimental inaccuracy of the mixing matrix elements of the so far 
observed quarks (the inaccuracy  of masses do not influence the results very much)  
is tough.}. 
It turns out that the experimental~\cite{datanew}
 inaccuracies are for the mixing matrices too large to tell 
trustworthy mass intervals for the quarks masses of the fourth family 
members~\footnote{We have not 
yet succeeded to repeat the calculations presented in Refs.~\cite{gn2014} with 
the newest data from Ref.~\cite{PDG2020}. Let us say that the accuracy of the
mixing matrix even for quarks remains  in Ref.~\cite{PDG2020} far 
from needed to predict the masses of the fourth two quarks. For the chosen 
masses of the four family quarks the mixing  matrix elements are expected to  
slightly change in the direction proposed by Eq.~(\ref{vudoldexp}).}. 
Taking into account the calculations of Ref.~\cite{datanew} 
fitting the experimental data (and the meson decays evaluations in the 
literature  as well as our own evaluations) the authors of the 
paper~\cite{gn2014} very roughly  estimate that the fourth family quarks 
masses might be pretty above $1$ TeV. 

Since the matrix elements of the $3 \times 3$ submatrix of the $4 \times 4$ 
mixing matrix depend weakly on the fourth family masses, the 
calculated mixing matrix
% (from the experimental data under the assumption that the mass matrices 
% manifest the symmetry of Eq.~(\ref{M0}))
offers the prediction to what values will more accurate measurements move
the present experimental data and also the fourth family mixing matrix 
elements in dependence of the fourth family masses, Eq.~(\ref{vudoldexp}): \\
$V_{u d}$ will stay the same or will very slightly decrease;  
$V_{u b}$ and $V_{c s}$, will still lower;
$V_{t d}$ will lower, and $V_{t b}$  will lower; 
$V_{u s}$ will slightly increase; 
$V_{c d}$ will (after decreasing) slightly rise; 
$V_{c b}$ will still increase and $V_{t s}$ will (after decreasing) increase. 
%%%%%%%%%%%%%%%%%%%%%%%%%%%%
 
%* Make shorter or skip this paragraph\\
In Eq.~(\ref{vudoldexp}) the matrix elements of the $4\times 4$ 
mixing matrix for quarks are presented, obtained when the $4\times 4$ mass 
matrices respect the symmetry of Eq.~(\ref{M0}) while the parameters of the 
mass matrices are fitted to %the old ($exp_o$)  and
the  ($exp$) experimental data~\cite{datanew}, Ref.~\cite{gn2014}. 
 The two choices of the fourth family quark masses are used in the calculations:
   $m_{u_4}= m_{d_4}=700$ GeV  ($scf_{1}$) and 
$m_{u_4}= m_{d_4}=$ $1\,200$ GeV ($scf_{2}$). 
In parentheses, $(\;)$ and $[\;\,]$, the  changes of the matrix elements are 
presented, which are due to the changes of the top mass
within the experimental inaccuracies: with the $m_{t} =$ $(172 + 3\times 0.76)$ 
GeV and $m_{t} =$ $(172 - 3\times 0.76)$, respectively  (if there are one, 
%number in parentheses only the last number is different, if there are 
two or more numbers in parentheses the last one or more numbers 
are different, if there is no parentheses no numbers are different) 
%Results are presented for two choices of $m_{u_{4}}$ $=m_{d_{4}}$, 
[arxiv:1412.5866].

%*CHECK PREDICTION WITH NEW \cite{PDG2920}

\vspace{2mm}

%\begin{tiny}
\begin{small}
\begin{equation}
\label{vudoldexp}
      |V_{(ud)}|= \begin{pmatrix}
    %
    % exp_o &   0.97425 \pm 0.00022    &  0.2252 \pm 0.0009   
%&  0.00415 \pm 0.00049 &    \\
     exp  &    0.97425 \pm 0.00022    &  0.2253 \pm 0.0008 
&  0.00413 \pm 0.00049&   \\
     \hline
    % old_1 &       0.97423                &  0.22531              &  0.00299  & 0.01021\\
    % old_2 &       0.97425                &  0.22536              &  0.00301  & 0.00474\\
     scf_1  &    0.97423(4)            &  0.22539(7)          
&  0.00299  &     0.00776(1)\\  
      scf_2  &    0.97423[5]            &  0.22538[42]        &  0.00299  
&  0.00793[466]\\ 
     \hline 
    % exp_o &   0.230   \pm 0.011      &  1.006  \pm 0.023     &  0.0409  \pm 0.0011&     \\
     exp  &  0.225   \pm 0.008      &  0.986  \pm 0.016     
&  0.0411  \pm 0.0013&   \\
    \hline
    %old_1  &   0.22526            &  0.97338              &  0.04238    & 0.00160 \\
    %old_2  &   0.22534            &  0.97336              &  0.04239    & 0.00212 \\
    scf_1  &  0.22534(3)       &  0.97335              &  0.04245(6) 
&   0.00349(60) \\  
    scf_2  &  0.22531[5]       &  0.97336[5]          &  0.04248     
&   0.00002[216] \\ 
    \hline
   % exp_o  &   0.0084  \pm 0.0006     &  0.0429 \pm 0.0026    &  0.89    \pm 0.07&      \\ 
    exp  &  0.0084  \pm 0.0006     &  0.0400 \pm 0.0027    
&  1.021   \pm 0.032&     \\
   \hline
    %old_1  &   0.00663                &  0.04197              &  0.99910   &0.00040 \\
   % old_2  &   0.00663                &  0.04198              &  0.99910   &0.00021\\
    scf_1  &  0.00667(6)            &  0.04203(4)         &  0.99909   
&     0.00038\\  
     scf_2  &  0.00667                &  0.04206[5]         &  0.99909   
&     0.00024[21] \\
   \hline
   %old_1    & 0.00959                  & 0.00388             & 0.00031    & 0.99995\\
   %old_2    & 0.00414                   & 0.00315             & 0.00011    &  0.99999\\
    scf_1   & 0.00677(60) & 0.00517(26)    & 0.00020    & 0.99996\\
   scf_2   & 0.00773      & 0.00178   & 0.00022  & 0.99997[9]
     \end{pmatrix}\,.
     \end{equation}
%\end{tiny}
\end{small}

\vspace{1mm}

Let me conclude that according to Ref.~\cite{gn2014} the masses of the  fourth 
family lie  much above the known three.
%, masses of quarks are close to each other.
%
The larger are  masses of the fourth family the larger are $V_{u_1 d_4}$
in comparison with $V_{u_1 d_3}$ and the more is valid that 
$V_{u_2 d_4} <V_{u_1 d_4}$, $V_{u_3 d_4}<V_{u_1 d_4}$. 
The flavour changing neutral currents are correspondingly weaker.

%One can see what\\
% B. Belfatto, R. Beradze, Z. Berezhiani, %"The CKM unitarity problem: 
%A trace of new physics at the TeV scale?"
% required in [arXiv:1906.02714v1], that \\
Let be noticed that the prediction of  Ref.~\cite{BBB}, 
$V_{u_1 d_4}> V_{u_1 d_3}$, $V_{u_2 d_4}<V_{u_1 d_4}$,  
$V_{u_3 d_4}<V_{u_1 d_4}$, agrees with the
prediction of Refs.~\cite{gn2014}.

In Ref.~\cite{NH2017newdata} the authors discuss the question why the existence
of the fourth family is not (at least yet) in contradiction with the 
experimental data.

\vspace{2mm}

{\bf ii.} The theory predicts the existence of several scalar fields. To the
lower four families two triplets and three singlets contribute, to the upper 
four families the same three singlets and different two triplets contribute,
Eq.~(\ref{commonAi}), Sects.~\ref{vectorscalar3+1},~\ref{scalar3+1}.
Some superposition of the three singlets and two triplets contributing to 
masses and to mixing matrices of quarks and leptons of the lower four 
families will be observed, representing so far the observed scalar higgs 
and Yukawa couplings.

\vspace{2mm}

{\bf iii.} The theory predicts the existence of besides the additional  
scalar fields also the additional vector gauge fields of very high mass, 
Sects.~\ref{vectorscalar3+1}, ~\ref{vector3+1}.

\vspace{2mm}

{\bf iv.} The theory predicts the existence of the upper four families of 
quarks and leptons and antiquarks and antileptons,  Table~\ref{Table III.}, 
with the same family members charges, Table~7 of Ref~\cite{nh2021RPPNP}, 
as are the charges of the lower four families, interacting correspondingly with 
the same vector gauge fields.  At low energies the upper four families are 
decoupled from the lower four families. 

The masses of the upper four families are determined by the two 
triplets ($ \vec{\tilde{A}}^{\tilde{2}}_{\stackrel{78}{(\pm)}}, 
   \vec{\tilde{A}}^{\tilde{N}_{\tilde{R}}}_{\stackrel{78}{(\pm)}}$) and 
three singlets ($A^{Q}_{\stackrel{78}{(\pm)}}, 
A^{Q'}_{\stackrel{78}{(\pm)}}, A^{Y'}_{\stackrel{78}{(\pm)}}$), the 
same singlets contribute also to masses of the lower four families,
Sect.~\ref{scalar3+1}.

 The stable of the upper four families offers the explanation 
for the appearance of the {\it dark matter} in our universe. 

Since the masses of the upper four families are much higher than the masses 
of the lower four families, the "nuclear" force among the baryons and mesons
of these quarks  and antiquarks differ a lot from the nuclear force of the 
baryons and fermions of the lower four families.

A rough estimation of properties of  baryons  of the stable fifth family 
members, of their behaviour  during the evolution of the universe and when 
scattering on the ordinary matter, as well as a study of possible limitations 
on the family properties due to the cosmological and direct experimental 
evidences are done in Ref.~\cite{gn2009}. 

In Ref.~\cite{nm2015} the weak and "nuclear"   scattering  of such very 
heavy baryons by ordinary nucleons is studied, showing that the 
cross section for such scattering is very small and therefore consistent 
with the observation of experiments so far, provided that the quark mass of this
 baryon is about 100 TeV or above.

In Ref.~\cite{gn2009} a simple hydrogen-like model is used to evaluate 
properties of baryons of these heavy quarks, with one gluon  exchange
determining the force among the constituents of the fifth family baryons~%
\footnote{The weak force and the electromagnetic force start to be at small 
distances due to heavy masses of quarks of the same order of 
magnitude as the colour force.}.

The authors of Ref.~\cite{gn2009} study the freeze out procedure of the 
fifth family quarks and antiquarks and the formation of  baryons and 
antibaryons up to the temperature  $ k_b T= 1$ GeV,  when the colour 
phase transition starts which depletes almost all the fifth family quarks 
and antiquarks,  while the colourless
% (neutral with respect to the colour and electromagnetic charge) 
fifth family neutrons with very small scattering cross section decouples 
long before (at $ k_b T= 100$ GeV).
%, Fig.~2 of Ref.~\cite{nhRPPNP},~\cite{gn2009}.
%% \ref{DiagramI.}.
%
%
%\begin{figure}[h]
% \begin{center}
% \includegraphics[width=15cm,angle=0]{slika.png}
% \caption{The dependence of the two number densities, $n_{q_5}$ of the fifth family
%  quarks and $n_{c_5}$ of the fifth family clusters of quarks, as functions of 
% $\frac{m_{q_5} \, c^2}{ k_b \, T}$ is presented 
% for the special values $m_{q_5} = 71 \,{\rm TeV}$. The estimated 
% scattering cross sections, entering into Boltzmann equation, are presented in
% Ref.~\cite{gn2009}, Eqs.~(2,3,4.5),
%  %$\eta_{c_5} = \frac{1}{50}$ and $\eta_{(q\bar{q})_b}=1$. 
% %We take $g^*=91.5$. 
% In the treated energy (temperature $ k_b T$) interval 
% the one gluon exchange gives the  main 
% contribution to the scattering cross sections % of Eq.(\ref{sigmasq}) 
% entering into the Boltzmann equations for 
% $n_{q_5}$ and $n_{c_5}$. %In the figure we make a choice of the  
% %parameters within the estimated  intervals.
% }
% \end{center}
% \label{DiagramI.}
% \end{figure}
%

The cosmological evolution 
suggests for the mass limits the range $10$ TeV $< m_{q_5}  < 
{\rm a \, few} \cdot 10^2$ TeV 
and for the  scattering cross sections 
$ 10^{-8}\, {\rm fm}^2\, < \sigma_{c_5}\, <   10^{-6} \,{\rm fm}^2  $. 
The measured density of  the  dark matter 
does not put much limitation on the properties of heavy enough clusters %
\footnote{   
In the case that the weak interaction determines the  cross section of  the neutron 
$n_5$, the interval for the  fifth family quarks would be 
$10\; {\rm TeV} < m_{q_5} \, c^2< 10^5$ TeV. }.

%\vspace{2mm}

%*TU 07.04. Pay attention on the new cosmological density of Paolo Salucci!! 
%But the sentence is already written, only the ref. is missed! 

\vspace{2mm}

The DAMA/LIBRA experiments~\cite{RitaB} limit, provided that they measure 
the heavy fifth family clusters, the quark mass in the interval:
 $ 200 \,{\rm TeV} < m_{q_{5}} < 10^5\, {\rm TeV}$, Ref.~\cite{gn2009}.

Baryons of the   fifth family are heavy, forming small 
enough  clusters with small enough scattering amplitude among themselves 
and with the ordinary matter to be the candidate for the dark matter.  

Masses of the stable fifth family of quarks and leptons are much above 
the fourth family members.

Although the upper four families carry the weak (of two kinds) and the 
colour charge, these group of four families are completely decoupled from 
the lower four families up to the $<10^{16}$  GeV when the breaks 
of symmetries are expected to recover.

%%%%%%%%%%%%%%%%%%%%%

\vspace{2mm}

\section{Conclusions}
\label{conclusions} 

The {\it spin-charge-family}  theory~\cite{norma93,norma95,norma2001,%
pikan2006,IARD2016,prd2018,n2019PIPII,nh2021RPPNP} assumes in 
$d=(13+1)$-dimensional space a simple
action, Eq.~(\ref{wholeaction}), for the massless fermions  and for the massless 
vielbeins and the two kinds of spin connection fields, with which fermions
 interact. 
 The description of the internal space of fermions with ''basis vectors''
 which are superposition of an odd products of the Clifford algebra objects and 
 of bosons with ''basis vectors''  which are superposition of  an even products 
 of the Clifford algebra objects  offers the explanation for spins,
charges and families of fermions and  their vector and scalar gauge fields,
as required by the {\it standard model}, while % Sect.~\ref{internalspace}, 
explaining as well the second quantization postulates for fermions and bosons.

Some of the predictions of the {\it spin-charge-family} theory can experiments 
soon confirm and correspondingly confirm (or reject) the theory. 
Because the theory offers meaningful answers to many open questions in 
 physics of elementary fermion and boson fields and in cosmology and because 
 the theory offers  more and more answers the more effort and work is put 
 into it,   it might very well be that the theory does offer the right next step 
 beyond the {\it standard model}.

The description of fermions and bosons, both second quantized, with the 
Clifford odd and the Clifford even ''basis vectors'', respectively, clarifies how 
strongly are all the properties of elementary fields determined by the internal 
space of fields, and that the internal space of fermions not only unifies spin, 
handedness, all the charges and families of fermions but manifests as well 
the strong connections with the corresponding boson vector and scalar
gauge fields.

The theory obviously needs more collaborators as it is necessary to find 
answers to questions, like:\\

{\bf i.} What is the dimension of space time? 
In any dimension $d=2 (2n+1)$ there namely exist fermions of only one
handedness, as discussed in Ref.~\cite{norma2021SQFB}, while in any 
subspace of this space there are fermions of both handedness.
{\bf i.a.} How can we look for anomalies of Kaluza-Klein theories in
higher dimensions?  {\bf i.b.} As well as for the renormalizability?

{\bf ii.} The spontaneous breaks of symmetries, from the starting one to the 
final ones, must carefully be done. {\bf ii.a.} The breaks from any $d=2 (2n+1)$
in steps to the observable $d=(3+1)$ must be done, following the number of 
massless families of fermions and the appearance of the vector and scalar gauge 
fields in each step. So far we studied only the breaks of symmetry for the toy models~\cite{NHD,ND012,familiesNDproc}, starting with $d=(5+1)$.
{\bf ii.b.} To learn more the electroweak break with the scalar fields defined in 
$d=2(2n+1), n=3,$ with the space index $(7,8)
$~\cite{NHD,ND012,familiesNDproc,TDN}  needs additional treating.

{\bf iii.} The second quantization 
of fermion and boson fields with the description of the internal space of fermions 
and bosons by  the Clifford  odd and even ''basis vectors'', respectively, is opening
a new insight in to quantum field theory. Ref.~\cite{norma2021SQFB} presents
only the first step to the second quantization of bosons by  the Clifford even ''basis 
vectors''. A further study is needed.

{\bf iv.} One irreducible representation of the Lorentz group in the internal space 
of fermions, Table~7 in Ref.~\cite{nh2021RPPNP} and Table~5 in 
Ref.~\cite{norma2021SQFB}, includes all the quarks and leptons and antiquarks
and antileptons observed so far (with not yet observed the right handed neutrinos 
and the left handed antineutrinos included). No Dirac sea is needed. {\bf iv.a.}
Additional  studies of masses of fermions and antifermions  in addition to those 
of Refs.~\cite{gn2014,AN2018} are needed.
%%%%%%%%%%%%%%%%11.11.2021 at 14:00
 
{\bf v.} So far only three families of quarks and leptons have been observed.
The {\it spin-charge-family} theory predicts the fourth family to the observed
three, very weakly coupled to the observed three with masses a few TeV or
higher. Although the accurately known $3\times 3$ submatrix of the 
$4\times 4$ unitary matrix determines the $4\times 4$ matrix uniquely, even 
the quarks mixing matrix is known far non accurately enough to enable  prediction
of masses of the fourth family, Ref.~\cite{gn2014}. {\bf v.a.} A further study of 
the properties of the $4\times 4$ mixing matrix  as following from the mass matrices 
of quarks and leptons with the known symmetries (what reduces the number of
free parameters to be fitted to the experimental data) is  needed and the way 
of improving the experimental accuracy needs to be suggested.
{\bf v.b.}The proof that the symmetry of mass matrices $\widetilde{SU}(2)
\times  \widetilde{SU}(2) \times U(1)$ keeps in all orders of loop corrections,
presented in Ref.~\cite{AN2018}, must be checked.
% (AN, paper is written, A must write a computer program.) 

{\bf vi.}  There are scalar fields which are colour triplets and antitriplets, predicted 
by the {\it spin-charge-family} theory~\cite{n2014matterantimatter}, which 
transform antileptons into quarks and antiquarks into quarks and back, causing in 
the expanding universe matter-antimatter asymmetry. The study is needed to see
their influence on the lepton number non conservation. 

{\bf vii.} A study of the coupling constants of fermions to the corresponding gauge 
vector and scalar fields in comparison with those of $SO(10)$ and $SO(13+1)$ is 
needed. 

{\bf viii.} The masses of the upper four families after the electroweak break and 
the influence of the neutrino condensate on their masses must be studied. 
{\bf viii.a.} The behaviour of the stable fifth family members, their ''freezing'' out 
and formation of neutral objects, interacting with the weak force, is needed and
their contribution to the {\it dark matter}. 
{\bf viii.b.} As well as the contribution of the heavy neutrinos to the x{\it dark 
matter}.

{\bf ix.} If the {\it spin-charge-family} theory is the right next step beyond the {\it standard
model}, it is worthwhile to find out what it has in common with all the theories and 
models which seems to be promising.

{\bf x.} And many more.

%%%%%%%%%%%%%%%%%
\appendix

\section{Infinitesimal generators of subgroups of $SO(13,1) group$}
\label{generators}
\begin{small}

The relations are taken from Ref.~\cite{nh2021RPPNP}.

The reader can calculate all the quantum numbers  of Table~5 in 
Ref.~\cite{norma2021SQFB} %\ref{Table so13+1.} 
and of Table~\ref{Table III.}, if taking into account
the generators of the two $SU(2)$  ($\subset SO(3,1)$ $\subset SO(7,1) \subset SO(13,1)$) groups, describing  spins and handedness of fermions, their two kinds of the weak charges, the
colour charges, the ''fermion'' charge,  as well as the family quantum numbers. 

One needs
\begin{eqnarray}
\label{so1+3}
&&\vec{N}_{\pm}(= \vec{N}_{(L,R)}): = \,\frac{1}{2} (S^{23}\pm i S^{01},
S^{31}\pm i S^{02}, S^{12}\pm i S^{03} )\,,\;\,
% \nonumber\\ &&N^{\pm}_{+}      = N^{1}_{+} \pm i \,N^{2}_{+} = 
%  - \stackrel{03}{(\mp i)} \stackrel{12}{(\pm )}\,, \quad N^{\pm}_{-}= 
% N^{1}_{-} \pm  i\,N^{2}_{-} =    \stackrel{03}{(\pm i)} \stackrel{12}{(\pm )}\,.
\vec{\tilde{N}}_{\pm}(=\vec{\tilde{N}}_{(L,R)}): =
 \,\frac{1}{2} (\tilde{S}^{23}\pm i \tilde{S}^{01}\,,
%\tilde{S}^{31}\pm i \tilde{S}^{02}, \tilde{S}^{12}\pm i \tilde{S}^{03} )\,,
\end{eqnarray}
the generators of the two $SU(2)$ ($SU(2)$ $\subset SO(4)$ $\subset SO(7,1) 
\subset SO(13,1)$) groups, describing  the weak charge, $\vec{\tau}^{1}$, and
the second kind of the weak charge, $\vec{\tau}^{2}$,  of fermions and 
the corresponding family quantum numbers
% in the {\it spin-charge-family } theory
%
 \begin{eqnarray}
 \label{so42}
 \vec{\tau}^{1}:&=&\frac{1}{2} (S^{58}-  S^{67}, \,S^{57} + S^{68}, \,S^{56}-  S^{78} )\,,%\nonumber\\
 \;\;
 \vec{\tau}^{2}:= \frac{1}{2} (S^{58}+  S^{67}, \,S^{57} - S^{68}, \,S^{56}+  S^{78} )\,,
 \nonumber\\
% \tau^{1\pm}         &=& (\mp)\, \stackrel{56}{(\pm )} \stackrel{78}{(\mp )} \,, \qquad   
% \qquad \qquad  \qquad \qquad \;\;
% \tau^{2\mp}=            (\mp)\, \stackrel{56}{(\mp )} \stackrel{78}{(\mp )} \,,
 \vec{\tilde{\tau}}^{1}:&=&\frac{1}{2} (\tilde{S}^{58}-  \tilde{S}^{67}, \,\tilde{S}^{57} + 
 \tilde{S}^{68}, \,\tilde{S}^{56}-  \tilde{S}^{78} )\,, \;\;
 \vec{\tilde{\tau}}^{2}:=\frac{1}{2} (\tilde{S}^{58}+  \tilde{S}^{67}, \,\tilde{S}^{57} - 
 \tilde{S}^{68}, \,\tilde{S}^{56}+  \tilde{S}^{78} ),\,\,\;\;
 \end{eqnarray}
and the generators of $SU(3)$ and $U(1)$ subgroups of $SO(6)$ $\subset SO(13,1)$, describing  the colour charge and the ''fermion'' charge  of fermions as well as the corresponding 
family quantum number $\tilde{\tau}^4$
%in the {\it spin-charge-family } theory
%
 \begin{eqnarray}
 \label{so64}
 \vec{\tau}^{3}: = &&\frac{1}{2} \,\{  S^{9\;12} - S^{10\;11} \,,
  S^{9\;11} + S^{10\;12} ,\, S^{9\;10} - S^{11\;12}\, ,  %\nonumber\\
  S^{9\;14} -  S^{10\;13} ,\,  \nonumber\\
  && S^{9\;13} + S^{10\;14} \,,  S^{11\;14} -  S^{12\;13}\,, %\nonumber\\
  S^{11\;13} +  S^{12\;14} ,\,  \frac{1}{\sqrt{3}} ( S^{9\;10} + S^{11\;12} - 
 2 S^{13\;14})\}\,,\nonumber\\
 \tau^{4}: = &&-\frac{1}{3}(S^{9\;10} + S^{11\;12} + S^{13\;14})\,,\;\;\nonumber\\
 \tilde{\tau}^{4}: = &&-\frac{1}{3}(\tilde{S}^{9\;10} + \tilde{S}^{11\;12} + \tilde{S}^{13\;14})\,.
 \end{eqnarray}
The (chosen) Cartan subalgebra operators, determining the commuting operators in the 
above equations,
is presented in Eq.~(\ref{cartanM}). 

The  hypercharge $Y$ and the electromagnetic charge $Q$ and the corresponding family
 quantum numbers then follows as
 \begin{eqnarray}
 \label{YQY'Q'andtilde}
 Y:= \tau^{4} + \tau^{23}\,,\;\; Q: =  \tau^{13} + Y\,,\;\; 
 Y':= -\tau^{4}\tan^2\vartheta_2 + \tau^{23}\,, %\;\;
 \;\; Q':= -Y \tan^2\vartheta_1 + \tau^{13} \,,&&,\nonumber\\
  \tilde{Y}:= \tilde{\tau}^{4} + \tilde{\tau}^{23}\,,\,\;\tilde{Q}:= 
  \tilde{Y} + \tilde{\tau}^{13}\,,\;\;
   \tilde{Y'}:= -\tilde{\tau}^{4} 
  \tan^2 \vartheta_2 + \tilde{\tau}^{23}\,,\;
  \;\; \tilde{Q'}= -\tilde{Y} \tan^2 \vartheta_1 
  + \tilde{\tau}^{13}\,. &&\,.
 % \,,%\nonumber\\
%%  %{\cal Y}:&=& {\cal \tau}^{4} + {\cal \tau}^{23}\,,\;\; {\cal Y'}:=
%%% -{\cal \tau}^{4}\tan^2 \vartheta_2 + 
 %% %{\cal \tau}^{23}\,,\;\;
 %% %{\cal Q}: =  {\cal \tau}^{13} + {\cal Y}\,,\;\; {\cal Q'}:=
 %%% -{\cal Y} \tan^2 \vartheta_1 + %\cal \tau}^{13}\,
 %% %\nonumber\\
  \end{eqnarray}
 %
% are presented.
% \begin{small}
Below are some of the above expressions written in terms of  nilpotents and projectors
 \begin{eqnarray}
\label{plusminus}
 N^{\pm}_{+}         &=& N^{1}_{+} \pm i \,N^{2}_{+} = 
 - \stackrel{03}{(\mp i)} \stackrel{12}{(\pm )}\,, \quad N^{\pm}_{-}= 
 N^{1}_{-} \pm  i\,N^{2}_{-} = 
   \stackrel{03}{(\pm i)} \stackrel{12}{(\pm )}\,,
%   N^{\pm}_{+}      &=& N^{1}_{+} \pm i \,N^{2}_{+} = 
 % - \stackrel{03}{(\mp i)} \stackrel{12}{(\pm )}\,, \quad N^{\pm}_{-}= 
 % N^{1}_{-} \pm  i\,N^{2}_{-} = 
%   \stackrel{03}{(\pm i)} \stackrel{12}{(\pm )}\,,
\nonumber\\
 \tilde{N}^{\pm}_{+} &=& - \stackrel{03}{\tilde{(\mp i)}} 
 \stackrel{12}{\tilde{(\pm )}}\,, \quad 
 \tilde{N}^{\pm}_{-}= %\tilde{N}^{1}_{-} \pm i\,\tilde{N}^{2}_{-} = 
   \stackrel{03} {\tilde{(\pm i)}} \stackrel{12} {\tilde{(\pm )}}\,,\nonumber\\ 
 \tau^{1\pm}         &=& (\mp)\, \stackrel{56}{(\pm )} \stackrel{78}{(\mp )} \,, \quad   
 \tau^{2\mp}=            (\mp)\, \stackrel{56}{(\mp )} \stackrel{78}{(\mp )} \,,\nonumber\\ 
 \tilde{\tau}^{1\pm} &=& (\mp)\, \stackrel{56}{\tilde{(\pm )}} 
 \stackrel{78}{\tilde{(\mp )}}\,,\quad   
 \tilde{\tau}^{2\mp}= (\mp)\, \stackrel{56}{\tilde{(\mp )}} \stackrel{78}{\tilde{(\mp )}}\,.
 \end{eqnarray}
\end{small}

%\appendix

\section*{Acknowledgment}
The author  thanks Department of Physics, FMF, University of Ljubljana, Society of 
Mathematicians, Physicists and Astronomers of Slovenia,  for supporting the research on the 
{\it spin-charge-family} theory by offering the room and computer facilities and Matja\v z 
Breskvar of Beyond Semiconductor for donations, in particular for the annual workshops entitled "What comes beyond the standard models". Thanks also to all the participants of the
Annual workshops ''What comes beyond the standard models'' for helpful discussions.

{}
\end{document}